\documentclass[a4paper,11pt,prd,noshowkeys,notitlepage,nofootinbib]{revtex4}
\pdfoutput=1  
\usepackage[latin1]{inputenc}
\usepackage{amsmath,amssymb,mathrsfs}
\usepackage{ae,aecompl}  
\usepackage{graphicx}
\usepackage[]{units}
\usepackage[usenames,dvipsnames]{color} 
\usepackage{hyperref}
\hypersetup{
   colorlinks=true,       
    linkcolor=Blue,          
    citecolor=Plum,        
    filecolor=magenta,      
    urlcolor=YellowOrange           
}
\usepackage{setspace}  
\usepackage{multirow} 
\providecommand{\tm}{\tilde{m}}
\providecommand{\tX}{\tilde{X}}

\providecommand{\Rz}{R^{(0)}}
\providecommand{\Uz}{U^{(0)}}

\providecommand{\hM}{\hat{M}}
\providecommand{\eps}{\epsilon}

\providecommand{\cH}{\mathcal{H}}

\providecommand{\om}{\omega}
\providecommand{\cp}{\mathsf{CP}}

\providecommand{\cpmutau}{\cp^{\mu\tau}}
\providecommand{\tm}{\tilde{m}}

\providecommand{\tp}{{\mss{\mathsf{T}}}}
\providecommand{\mss}[1]{\mbox{\scriptsize $#1$}}
\providecommand{\ml}[1]{\mbox{\large $#1$}}
\providecommand{\eq}[1]{\begin{equation} #1 \end{equation}}
\providecommand{\eqali}[1]{\begin{equation}\begin{aligned} #1
    \end{aligned}\end{equation}}

\DeclareMathOperator{\re}{\mathrm{Re}} 
\DeclareMathOperator{\diag}{\mathrm{diag}} 
\providecommand{\mtrx}[1]{\begin{pmatrix} #1 \end{pmatrix}}
\providecommand{\ums}[2][1]{\ml{\tfrac{#1}{#2}}} 

\usepackage{mathtools}
\DeclarePairedDelimiter\norm{\lVert}{\rVert}%

\usepackage{bbm}
\providecommand{\id}{{\mathbbm{1}}} 
\providecommand{\xlink}[1]
  {\href{http://arxiv.org/abs/#1}{arXiv:#1}}

\begin{document}
\title{
Leptogenesis in the $\mu\tau$ basis
}
\author{C.~C.~Nishi}
\email{celso.nishi@ufabc.edu.br}
\affiliation{
Centro de Matemática, Computação e Cognição\\
Universidade Federal do ABC, 09.210-170,
Santo André, SP, Brazil
}
\author{Chee Sheng Fong}%
\email{sheng.fong@ufabc.edu.br}
\affiliation{ 
Centro de Ci\^encias Naturais e Humanas\;\;\\
Universidade Federal do ABC, 09.210-170,
Santo André, SP, Brazil
}

\begin{abstract} 
We formulate three-flavor type-I leptogenesis in the $\mu\tau$ basis which is convenient because in the three-flavor regime, both $\mu$ and $\tau$ charged lepton Yukawa interactions are in thermal equilibrium and the thermal bath is symmetric under the exchange $\mu \leftrightarrow \tau$. We apply this formalism to models with $\mu\tau$-reflection $\cpmutau$ symmetry. We confirm the previous result that leptogenesis fails in the three-flavor regime with exact $\cpmutau$ symmetry. Allowing $\cpmutau$ symmetry to be broken to various degrees, we show that leptogenesis can be successful in the three-flavor regime only in certain tuned parameter space, which could further imply additional symmetry is at play. As a bonus, we derive analytical expressions which could be utilized whenever the branching ratios for the decays to $\mu$ and $\tau$ flavors are equal or approximately so.		
\end{abstract}
\maketitle
\section{Introduction}

The leptogenesis mechanism for explaining the observed asymmetry between matter and antimatter in the universe is a beautiful and economical byproduct of the seesaw explanation for the smallness of neutrino masses\,\cite{leptog}.
The heavy degrees of freedom that suppress neutrino masses are also the ones that decay violating CP and induce the necessary lepton number asymmetry that are converted to the observed baryon asymmetry.

The typical mass scale for these heavy degrees of freedom is $M\sim 10^{14}\,\unit{GeV}$ if they contribute at tree level to the light neutrino masses and if their couplings to the SM leptons are order one.
For the simplest type I seesaw, these heavy degrees of freedom are SM singlets, commonly called heavy right-handed neutrinos $N_{i}$.
For this simple case and for the high scale of $10^{14}\,\unit{GeV}$ or above, leptogenesis operates in a regime where the flavor content of the generated lepton asymmetry is not distinguishable, a regime known as the one-flavor regime.

Of course, without new states at intermediate scales, such a heavy right-handed neutrinos are not observable in terrestrial experiments and are very difficult to probe by other means.
Lowering the scale of these right-handed neutrinos would be desirable to increase observability.
However, the Davidson-Ibarra bound\,\cite{Davidson:2002qv} constrains the maximum amount of total CP asymmetry that can be generated and consequently sets a lower limit of around $10^9\,\unit{GeV}$ for the leptogenesis scale induced by the total CP asymmetry\,\cite{Buchmuller:2004nz}.
The bound is not applicable if flavor effects are at play: the Yukawa interactions of the SM may be fast enough to distinguish some charged lepton flavors so that the asymmetry accummulated in some flavors may follow a different dynamics\,\cite{Nardi:2006fx,Abada:2006fw,Abada:2006ea}.
In particular, it becomes possible to go below $10^{9}\,\unit{GeV}$ where leptogenesis operates in the three flavor regime where all $e,\mu,\tau$ flavors can be distinguished in the plasma\,\cite{Garbrecht:2013urw,Garbrecht:2014kda}.
Without resorting to resonant enhancement of CP violation\,\cite{Pilaftsis:1997jf,Pilaftsis:2003gt} but only with flavor effects, some form of fine-tuning is still unavoidable if one were to go much below $10^9$ GeV\,\cite{pascoli.turner}.

Although successful in explaining the lightness of neutrino masses, the seesaw mechanism by itself cannot explain the pattern of large mixing angles and the relative scale of neutrino masses that was uncovered in the last decades\,\cite{nobel}.
One approach to increase predictive power is to assume a flavor symmetry acting on the horizontal space of the three families of lepton fields; see Refs.\,\cite{GF:review} for a review.
Unfortunately, if all the mixing angles are fixed by symmetry, one can show that there is no CP violation at low energy\,\cite{grimus.fonseca}.
However, neutrino oscillation experiments are almost excluding CP conserving values for the Dirac CP phase\,\cite{global.fit,t2k}, one of the remaining low energy parameters yet to be measured in the lepton sector.

One of the simplest symmetries capable of predicting all the CP phases at low energy, yet allowing for CP violation, is a symmetry called $\mu\tau$-reflecion or $\cpmutau$ in which 
$\nu_\mu$ and $\nu_\tau$ interchange is combined with CP symmetry\,\cite{mutau-r:HS,mutau-r:GL}; see also \cite{cp.mutau,mutau-r:review}.
This symmetry predicts a maximal Dirac CP phase $\delta=\pm 90^\circ$ and trivial Majorana phases together with maximal atmospheric angle.
The maximal values for the atmospheric angle and maximal $\delta_\cp=-\pi/2$ agrees with current global fits \cite{global.fit} and it is strenghened by the recent T2K result \cite{t2k}.

Leptogenesis in the presence of the $\cpmutau$ symmetry has been studied in the past and it was shown that leptogenesis was not successful in the one\,\cite{mutau-r:GL} and three flavor\,\cite{cp.mutau} regimes.
The first failure is due to the vanishing of the total CP asymmetry while the second one follows because the flavor projectors are $\mu\tau$ symmetric and the washout rates in the $\mu$ and $\tau$ flavors are the same. 
In the two flavor regime, the necessary lepton asymmetry can be easily generated\,\cite{cp.mutau}, even in a highly predictive scenario where a texture zero is present in the heavy right-handed neutrino mass matrix\,\cite{cpmutau:high}.

Here we will analyze leptogenesis with $\cpmutau$ symmetry in various temperature regimes. 
Firstly, in Sec.\,\ref{sec:mutau.basis},
we formulate Boltzmann equations for three-flavor leptogenesis in a basis we call the $\mu\tau$ basis where we explore the symmetry of the leptogenesis dynamics under $\mu\tau$ relabeling.
If, in addition, the underlying theory is invariant by some form of $\mu\tau$ symmetry such as $\cpmutau$, the Boltzmann equations decouple into two distinct pieces which evolves independently.
In this case, we confirm in Sec.\,\ref{sec:exact_mutau} the known result of the failure of three-flavor leptogenesis in the presence of $\cpmutau$ considering the flavor effects in full generality.
The question that follows is then the amount of symmetry breaking necessary for successful three-flavor leptogenesis. The amount of symmetry breaking may not necessarily be small because $\cpmutau$ may be generalized to include nontrivial Majorana phases\,\cite{real.cpmutau,king.nishi}.
One example can be seen in Ref.\,\cite{king.zhou} which is a more specific version of the 
Littlest seesaw model\,\cite{LSS}.

The rest of this paper is organized as follows.
Sec.\,\ref{sec:mutau_breaking} analyzes the necessary amount of $\cpmutau$ breaking that is necessary for successful leptogenesis. 
We conclude that three-flavor leptogenesis is barely possible with small breaking.
So Sec.\,\ref{sec:large.breaking} considers large $\cpmutau$ breaking and 
demonstrates that some amount of fine-tuning is unavoidable.
Sec.\,\ref{mutau-U:models} shows some examples of models where large $\cpmutau$ breaking may occur only in the CP asymmetries but not on the flavor projectors. 
The conclusions can be seen in Sec.\,\ref{sec:concl} and the appendices contain auxiliary material.

\section{Three-flavor Leptogenesis in the $\ml{\mu\tau}$ basis}
\label{sec:mutau.basis}

The type-I seesaw Lagrangian in the basis where charged lepton Yukawa $y_\alpha$, $\alpha=e,\mu,\tau$, and Majorana mass $M_i$, $i=1,2,3,$ for the right-handed neutrinos $N_i$ are real and diagonal is given by
\eq{
-{\cal L} =
M_i \bar N_i N_i^c 
+ \lambda_{i\alpha} \bar N_{i} \tilde H^\dagger \ell_\alpha
+ y_\alpha \bar \ell_\alpha H e_{\alpha},
}
where $\ell_\alpha$ and $H$ are respectively the SM lepton and Higgs doublets with $\tilde H =i \sigma_2 H^*$, and $e_{\alpha}$ are the right-handed charged leptons.

In the three-flavor regime, $T \lesssim 10^9$ GeV, both $\mu$ and $\tau$ charged lepton Yukawa interactions are in thermal equilibrium,
and all three lepton flavors can be distinguished in the plasma.
Since $\mu$ and $\tau$ leptons are both massless and carry the same SM quantum numbers, the cosmic thermal bath is symmetric under the exchange $\mu \leftrightarrow \tau$, i.e., the ``dynamics'' with such a relabelling is the same.
In this case, it is convenient to describe leptogenesis in the $\mu\tau$ basis that we will develop here.

In this regime, the Boltzmann equations (BEs) for $\Delta_{\alpha}\equiv\frac{B}{3}-L_{\alpha}$ charge produced from the decay and inverse decays of right-handed neutrinos $N_i\leftrightarrow \ell_\alpha H$, $N_i\leftrightarrow \bar{\ell}_\alpha H^*$ are given by
\begin{eqnarray}
\frac{dY_{\Delta_{\alpha}}}{dz} & = & -\sum_{i}\left[\epsilon_{i\alpha}D_{i}\left(\frac{Y_{N_{i}}}{Y_{N_{i}}^{{\rm eq}}}-1\right)-\frac{1}{2}P_{i\alpha}D_{i}\left(\frac{Y_{\ell_\alpha}}{Y_{\ell_\alpha}^{{\rm eq}}}+\frac{Y_{H}}{Y_{H}^{{\rm eq}}}\right)\right]\nonumber \\
& = & -\sum_{i}\left[\epsilon_{i\alpha}D_{i}\left(\frac{Y_{N_{i}}}{Y_{N_{i}}^{{\rm eq}}}-1\right)-\frac{1}{2}P_{i\alpha}D_{i}\frac{1}{Y^{{\rm eq}}}\sum_{\beta}\left(A_{\alpha\beta}+C_{\beta}\right)Y_{\Delta_{\beta}}\right], \label{BE:3-f} \\
\frac{dY_{N_i}}{dz} & = &
- D_{i}\left(\frac{Y_{N_{i}}}{Y_{N_{i}}^{{\rm eq}}}-1\right), \label{BE:N}
\end{eqnarray}
where $z\equiv\frac{M_{1}}{T}$ with $M_i$ the mass of $N_i$, $Y^{\rm eq} = \frac{15}{8\pi^2 g_\star}$ with $g_\star$ the total relativistic degrees of freedom ($g_\star = 106.75$ for the SM), $Y_a \equiv \frac{n_a}{s}$ with $n_a$ the number density of particle $a$ and $s = \frac{2\pi^2}{45} g_\star T^3$ the cosmic entropic density,  and $Y_{N_i}^{\rm eq} = \frac{45}{2\pi^4 g_\star} a_i^2 z^2 {\cal K}_2 (a_i z)$ with $a_i \equiv M_i/M_1$ and  ${\cal K}_n$ the modified Bessel function of the second kind of order $n$. 
The flavored CP parameters are defined as~\cite{Covi:1996wh}
\begin{eqnarray}
\epsilon_{i\alpha} &\equiv& 
\frac{\Gamma(N_i \to \ell_\alpha H)-\Gamma(N_i \to \bar\ell_\alpha H^*)}{\Gamma_{N_i}} \nonumber \\
& = & \frac{1}{8\pi (\lambda \lambda^\dagger)_{ii}}
\sum_{j\neq i}
\left\{ 
{\rm Im} \left[(\lambda\lambda^\dagger)_{ij} \lambda_{i\alpha}\lambda_{j\alpha}^*\right]
g(x_{ji})
+ {\rm Im}\left[(\lambda\lambda^\dagger)_{ji} \lambda_{i\alpha}\lambda_{j\alpha}^*\right]
\frac{1}{1-x_{ji}}
\right\},
\label{eq:CPfla}
\end{eqnarray}
where $x_{ji} \equiv \frac{M_j^2}{M_i^2}$, $\Gamma_{N_i} = \frac{(\lambda\lambda^\dag)_{ii} M_i}{8\pi}$ is the $N_i$ tree-level total decay width and the one-loop function is given by
\eq{
g(x) = \sqrt{x}\left[\frac{1}{1-x} + 1 - (1+x)\ln\left(\frac{1+x}{x}\right)\right].
\label{eq:1loop_fun}
}
The decay reaction is described by 
\eq{
\label{eq:decay}
D_{i}\equiv Y_{N_{i}}^{{\rm eq}}\frac{\Gamma_{N_{i}}}{{\cal H}z}\frac{{\cal K}_{1}\left(a_i z\right)}{{\cal K}_{2}\left(a_i z\right)},
}
where ${\cal H}=1.66\sqrt{g_\star}\frac{T^2}{M_{\rm Pl}}$ is the Hubble rate with $M_{\rm Pl} = 1.22 \times 10^{19}\,{\rm GeV}$.
Finally, the flavor projectors are
\eq{
\label{flavor.proj}
	P_{i\alpha}\equiv \frac{|\lambda_{i\alpha}|^2}{(\lambda\lambda^\dag)_{ii}}\,,
}
with $\sum_\alpha P_{i\alpha}=1$. 

Since the system is symmetric under the exchange $\mu \leftrightarrow \tau$, the generic forms of the flavor coefficients $A\sim 3\times 3$ and $C\sim 1\times 3$ 
are\footnote{The numerics of the entries will change with temperature based on the reactions which are in or out of thermal equilibrium (see Appendix \ref{app:A_C}).}
\begin{eqnarray}
A & = & \left(\begin{array}{ccc}
A_{ee} & A_{e\mu} & A_{e\mu}\\
A_{\mu e} & A_{\mu\mu} & A_{\mu\tau}\\
A_{\mu e} & A_{\mu\tau} & A_{\mu\mu}
\end{array}\right),\\
C & = & \left(\begin{array}{ccc}
C_e & C_\mu & C_\mu\end{array}\right).
\end{eqnarray}
They satisfy
\eq{
	\label{eq:AC_sym}
	XAX=A\,,\quad
	CX=C\,,
}
where
\eq{
	X\equiv \mtrx{1 & 0&0\cr 0& 0&1\cr 0& 1&0}\,.
}
Defining the flavor matrix
\eq{
	F_{\alpha\beta} \equiv \left(A_{\alpha\beta}+C_{\beta}\right),
	\label{eq:matrix_F}
}
we can rewrite eq.~\eqref{BE:3-f} in vector notation as
\begin{eqnarray}
\frac{d\vec{Y}_{\Delta}}{dz} & = & -\sum_{i}\left[\vec{\epsilon}_{i}D_{i}\left(\frac{Y_{N_{i}}}{Y_{N_{i}}^{{\rm eq}}}-1\right)-\frac{1}{2Y^{{\rm eq}}}D_{i}
P_i F \vec{Y}_{\Delta}\right],
\label{eq:BE_matrix_form}
\end{eqnarray}
where $P_i = {\rm diag}(P_{ie},P_{i\mu},P_{i\tau})$, $\vec{\epsilon}_{i}=\left(\epsilon_{ie},\eps_{i\mu},\epsilon_{i\tau}\right)^{T}$ 
and $\vec{Y}_{\Delta}=\left(Y_{\Delta_{e}},Y_{\Delta_{\mu}},Y_{\Delta_{\tau}}\right)^{T}$.
Note that $P_i$ and $F$ are $3\times 3$ matrices.

Next, let us define the projectors into $\mu\tau$ even and odd subspace
\eq{
	X_\pm \equiv \frac{1}{2}(I_3 \pm X),
	\label{eq:mutauproj}
}
with $X_\pm^2 = X_\pm$ and $X_+ X_- = 0$. We can recast eq.~\eqref{eq:BE_matrix_form} in terms of $\mu\tau$ even and odd components $\vec Y_{\pm} \equiv X_\pm \vec Y_\Delta$ as
\begin{eqnarray}
\frac{d \vec{Y}_{+}}{dz} 
& = & -\sum_{i}\left[ X_+ \vec{\epsilon}_{i}D_{i}\left(\frac{Y_{N_{i}}}{Y_{N_{i}}^{{\rm eq}}}-1\right)
-\frac{1}{2Y^{{\rm eq}}}D_{i}
\left( X_+ P_i F \vec{Y}_+ 
+ X_+ P_i F  \vec{Y}_- \right)
\right], \label{eq:BE_matrix_form_p} \\
\frac{d \vec{Y}_{-}}{dz}
& = & -\sum_{i}\left[ X_- \vec{\epsilon}_{i}D_{i}\left(\frac{Y_{N_{i}}}{Y_{N_{i}}^{{\rm eq}}}-1\right)
-\frac{1}{2Y^{{\rm eq}}}D_{i}
\left(X_- P_i F  \vec{Y}_+ 
+ X_- P_i F  \vec{Y}_- \right)
\right].
\label{eq:BE_matrix_form_m}
\end{eqnarray}
Note that the $\mu\tau$ even component is two-dimensional $\vec Y_+ = \left(Y_{\Delta_e},(Y_{\Delta_\mu} + Y_{\Delta_\tau})/2,(Y_{\Delta_\mu} + Y_{\Delta_\tau})/2 \right)^T$ whereas the $\mu\tau$ odd component is one-dimensional $\vec Y_- = \left(0,(Y_{\Delta_\mu} - Y_{\Delta_\tau})/2,-(Y_{\Delta_\mu} - Y_{\Delta_\tau})/2\right)^T$.

We will further rewrite $P_i$ as
\eq{
	P_i \equiv \Sigma_i  + \delta_i,
}
where
\begin{eqnarray} 
\Sigma_i &\equiv& 
\diag\left(P_{ie},\ums{2}(P_{i\mu}+P_{i\tau}),\ums{2}(P_{i\mu}+P_{i\tau})\right)
,  \\
\delta_i &\equiv& 
\ums{2}(P_{i\mu}-P_{i\tau})\cdot
\diag(0,1,-1).
\end{eqnarray}
Using the fact that $X$ commutes with $\Sigma_i$ and anticommutes with $\delta_i$, i.e., $[X,\Sigma_i] = 0$ and $\{X,\delta_i\}$ = 0, we have
\begin{eqnarray}
[X_\pm,\Sigma_i] &=& 0, \\
X_\pm \delta_i &=& \delta_i X_\mp.
\end{eqnarray}
The above identities together with $[X_\pm,F]=0$, which follows from eq.~\eqref{eq:AC_sym}, simplify eqs.~\eqref{eq:BE_matrix_form_p} and \eqref{eq:BE_matrix_form_m} to
\begin{eqnarray}
\frac{d \vec{Y}_{+}}{dz} 
& = & -\sum_{i}\left[ X_+ \vec{\epsilon}_{i}D_{i}\left(\frac{Y_{N_{i}}}{Y_{N_{i}}^{{\rm eq}}}-1\right)
-\frac{1}{2Y^{{\rm eq}}}D_{i}
\left( \Sigma_i F \vec{Y}_+ 
+ \delta_i F \vec{Y}_- \right)
\right], \label{eq:BE_matrix_form_p2} \\
\frac{d \vec{Y}_{-}}{dz} 
& = & -\sum_{i}\left[ X_- \vec{\epsilon}_{i}D_{i}\left(\frac{Y_{N_{i}}}{Y_{N_{i}}^{{\rm eq}}}-1\right)
-\frac{1}{2Y^{{\rm eq}}}D_{i}
\left( \delta_i F \vec{Y}_+ 
+ \Sigma_i F \vec{Y}_- \right)
\right].
\label{eq:BE_matrix_form_m2}
\end{eqnarray}

Finally, the total $B-L$ asymmetry is given by
\eq{
Y_\Delta = \sum_p \left[ (Y_+)_p + (Y_-)_p \right] = \sum_p  (Y_+)_p ,
}
where $p$ refers to the vector component of $\vec Y_\pm$ and we have used $\sum_p  (Y_-)_p = 0$. Although the $\mu\tau$ odd component $\vec Y_-$ does not contribute directly to $Y_\Delta$, it does contribute to $\vec Y_+$ through the right-handed side of eq.~\eqref{eq:BE_matrix_form_p2}. However, if $P_{i\mu} = P_{i\tau}$, as when $\mu\tau$ interchange or $\mu\tau$-reflection symmetry is valid in the neutrino sector, it follows that $\delta_i = \mathbf{0}$ and the BEs for $\vec Y_+$ and $\vec Y_-$ in eqs.~\eqref{eq:BE_matrix_form_p2} and \eqref{eq:BE_matrix_form_m2} decouple. In this case, we only have to solve for $\vec Y_+$ in eq.~\eqref{eq:BE_matrix_form_p} which is two-dimensional, i.e., essentially a two-flavor scenario. 
Analytical approximate solutions to eqs.~\eqref{eq:BE_matrix_form_p2} and \eqref{eq:BE_matrix_form_m2} are derived in Appendix \ref{app:approx_sol_mutau} for the case of $\delta_i = \mathbf{0}$ and $|\delta_i| \ll \mathbf{1}$.

Eventually, as the ElectroWeak (EW) spharelons processes go out of equilibrium at $T \sim 130$ GeV~\cite{DOnofrio:2014rug}, the final baryon asymmetry is frozen to be~\cite{DOnofrio:2014rug,Harvey:1990qw,Inui:1993wv}
\begin{equation}
Y_B = \frac{30}{97} Y_\Delta,
\label{eq:YBYDelta}
\end{equation}
where we have assumed that the EW symmetry is already broken as suggested in~\cite{DOnofrio:2014rug} and do not include the contribution of top quark. In this work, we fix $Y_B|_{\rm exp} = 8.7 \times 10^{-11}$ as indicated by the Planck measurement~\cite{planck.2018}.

\section{Exact $\cpmutau$ limit}
\label{sec:exact_mutau}

It was shown in Ref.\,\cite{cp.mutau} that leptogenesis with $\mu\tau$-reflection or 
$\cpmutau$ can be successful only in the two-flavor regime where $10^{9}\,\unit{GeV}\lesssim M_1\sim T\lesssim 10^{12}\,\unit{GeV}$. 
Here we review and confirm this result from a more precise calculation that includes off-diagonal flavor effects\,\cite{Nardi:2006fx,Abada:2006fw,Abada:2006ea} which, in principle, potentially could (but do not) source flavor dependent lepton asymmetries in the three-flavor regime.
These effects were not considered in Ref.\,\cite{cp.mutau}.
Later, off-diagonal flavor effects were briefly considered in Ref.\,\cite{gen.mutau} but only a numerical example was given to illustrate the general case.
Here, we treat these flavor effects in full analytical generality and, additionally, identify the features that preclude successful leptogenesis.

In the limit where $\mu\tau$-reflection symmetry or $\cpmutau$ is exact in the neutrino sector, the flavored CP parameters in eq.~\eqref{eq:CPfla} 
satisfy\,\cite{cp.mutau}
\eq{
\label{eps}
\epsilon_{ie}=0 \text{~~and~~}\epsilon_{i\mu}=-\epsilon_{i\tau}\,.
}
The symmetry also relates the $N_i$ Yukawa couplings as
\eq{
\label{lambda:mutau}
\lambda^2_{i e}\text{~is real}\,,\quad
|\lambda_{i \mu}|=|\lambda_{i\tau}|\,,
}
which implies 
\eq{
\label{Pmutau}
	P_{i\mu} = P_{i\tau}.
}
If leptogenesis happens above $10^{12}$ GeV where lepton flavor effect
is not at play, it is clear that leptogenesis fails because\,\cite{mutau-r:GL,cp.mutau}
\eq{
\label{sum=0}
\epsilon_i \equiv \sum_{\alpha}\epsilon_{i\alpha}=0
\,.
}

Interestingly, while the total CP parameter $\epsilon_i$ is zero, the individual flavored CP parameters $\epsilon_{i\alpha}$ could be much larger than the Davidson-Ibarra bound \cite{Davidson:2002qv} on $\epsilon_i$ for hierarchical $N_i$.  This shows that one might be able to realize purely flavored leptogenesis\,\cite{Nardi:2006fx,AristizabalSierra:2009bh} below the one-flavor regime $T \lesssim 10^{12}$ GeV where interactions mediated by $\tau$ charged lepton Yukawa get into thermal equilibrium.\footnote{%
We assume the $\cpmutau$ symmetry for the charged lepton sector is broken at a scale higher than the leptogenesis scale so that Yukawa couplings $y_\alpha$ coincide with the SM ones. See Ref.\,\cite{cp.mutau} for some ways to implement it.}
Indeed, this was shown to be the case in Ref.\,\cite{cp.mutau} in the two-flavor regime $10^9\,{\rm GeV} \lesssim T \lesssim 10^{12}$ GeV. Focusing on diagonal flavor effects,  Ref.\,\cite{cp.mutau} also demonstrated that in the three-flavor regime $T \lesssim 10^9$ GeV, there is an exact cancellation resulting in vanishing final baryon asymmetry.

In the three-flavor regime, we can analyze the consequences of $\cpmutau$ on leptogenesis using the BEs in the $\mu\tau$ basis. 
Here, we keep the index $i$ to demonstrate that our analysis holds for any $N_i$ in the three-flavor regime.
Due to eq.~\eqref{eps}, we can see that $\vec{\eps}_i$ is odd under $\mu\tau$ interchange and as a result
\eq{
	\label{eq:CP_mutau}
	X_+ \vec{\eps}_i= 0,\;\;\;X_- \vec{\eps}_i=\vec{\eps}_i= (0,-\epsilon_{i\tau},\epsilon_{i\tau})^T,
}
As $\cpmutau$ symmetry further constrains $P_{i\mu}=P_{i\tau}$ from eq.~\eqref{Pmutau}, the BEs for $\vec Y_+$ and $\vec Y_-$ in eqs.~\eqref{eq:BE_matrix_form_p} and \eqref{eq:BE_matrix_form_m} decouple and we only need to solve for $\vec Y_+$. The solution for $\vec Y_+$ is:
\eq{
	\label{eq:3fla_sol}
	\vec{Y}_+(z) = \vec Y_+(z_0) 
	\exp \left[ \frac{1}{2Y^{{\rm eq}}} \int_{z_0}^{z} dz' \sum_i \Sigma_i F D_i(z') \right]\,.
}
where $z_0 = M_i/T_0$ with $T_0$ the initial temperature. In the absence of preexisting asymmetry, $\vec Y_+(z_0) = 0$, and the final $B-L$ asymmetry remains zero.
The ``preexisting'' asymmetry can also come from two-flavor leptogenesis due to decays of heavier $N_i$.
In this case, one needs to make sure that the asymmetry survives the three-flavor washout from the lightest $N_1$. 
For large $K_1$ and generic $P_{1\tau}$, one needs large preexisting asymmetry to survive washout. 
For special parameters, such as for $P_{1\tau}\approx 0$ or $P_{1\tau}\approx 0.5$, $N_1$ washout can be very weak, even for large $K_1$.
In the absence of special conditions which allow the survival of a preexisting asymmetry, we have proven that three-flavor leptogenesis fails in the limit of exact $\cpmutau$.
In the context of $\cpmutau$, a detailed study of the contribution from $N_2$ leptogenesis\,\cite{N2.lepto} is considered in Ref.\,\cite{samanta.sen} and will not be considered further in this work.

In between $10^{9}\,{\rm GeV}\lesssim T\lesssim10^{12}\,{\rm GeV}$
where only the $\tau$-lepton flavor can be distinguished, leptogenesis can
proceed in the two-flavor regime that distinguishes $\tau$ from $e+\mu$. In this case, the
BEs are the same as \eqref{BE:3-f} but now the flavor indices run through $\alpha=e+\mu,\tau$, with 
\eq{
Y_{\Delta_{e+\mu}}=Y_{\Delta_{e}}+Y_{\Delta_{\mu}}\,,
\quad
P_{i,e+\mu}=P_{ie}+P_{i\mu}\,,\quad
\eps_{i,e+\mu}=\eps_{ie}+\eps_{i\mu}\,.
}
So the flavor coefficients $A,C$ have sizes $2\times 2$ and $1\times 2$ respectively.
We can put the equations in matricial form as in \eqref{eq:BE_matrix_form} with the respective modifications that include $\vec{Y}_{\Delta}=(Y_{\Delta_{e+\mu}},Y_{\Delta_{\tau}})$ and $\vec{\eps_i}=(\eps_{i,e+\mu},\eps_{i\tau})$.

With $\cpmutau$, we can still define an effective $\mu\tau$ interchange:
\eq{
\tX =\mtrx{0&1\cr 1&0}\,.
}
Under this interchange, we still have odd CP parameter,
\eq{
\tX \vec{\eps}_i=-\eps_i\,,
}
but $A,C$ are not symmetric by interchange and $P_{e+\mu}\neq P_\tau$ as well.
So the $F$ matrix is also not $\mu\tau$ symmetric.
Therefore, the BEs for the $\mu\tau$ even and odd components no longer decouple and an asymmetry in the total lepton number direction generically survives and may reach the necessary value\,\cite{cp.mutau,cpmutau:high}.

\section{{\protect\large$\mu-\tau$} symmetry breaking}
\label{sec:mutau_breaking}

Having confirmed that leptogenesis with $\cpmutau$ symmetry cannot proceed successfully in the one- or three-flavor regimes, we will analyze here how much breaking of $\cpmutau$ is necessary to account for successful leptogenesis in these regimes. We initially focus on small breakings of $\cpmutau$ such as induced by RGE running\,\cite{mutau-r:review,nath}.
But we should keep in mind that models with large breakings solely in the Majorana phases can be constructed naturally\,\cite{real.cpmutau,king.nishi,king.zhou}.
In Ref.\,\cite{king.nishi}, $\cpmutau$ symmetry was denoted as $\mu\tau$-R and the generalized version when Majorana phases were generic was called $\mu\tau$-U.
In other words, $\mu\tau$-U ensures $\theta_{23}=45^\circ$ and $\delta=\pm 90^\circ$ but free Majorana phases.
We will use the nomenclature $\mu\tau$-R and $\mu\tau$-U when the need to distinguish arises. 

For simplicity, we consider the scenario where leptogenesis proceeds only through decays of $N_1$ and drop the corresponding subscript ``1'' unless it is required for clarity. This is realized for instance when the reheating temperature $T_R$ falls in the range $M_1 <T_R \ll M_2 < M_3$. If $M_1$ and $M_2$ are of similar order but not quasi-degenerate (no resonant enhancement), both will contribute constructively, resulting in an enhancement of about a factor of two in the weak washout regime. In the strong washout regime, there is no enhancement since additional contribution in the source term is compensated by the additional washout.

\subsection{One-flavor regime}
\label{sec:1f:break}

Here we consider the scenario where $10^{12}\,{\rm GeV} \lesssim M_1 < T_R$ such that $N_1$ leptogenesis occurs in the one-flavor regime.
We assume leptogenesis can be sourced by a small breaking of $\cpmutau$ leading to
\eq{
\label{total.eps:break}
\epsilon \equiv \eps_e+\eps_\mu+\eps_\tau=\delta\eps\,,
}
where
\eq{
	\label{eq:small_break}
	|\delta\eps|\ll |\eps_\tau|\,.
}
Such a small breaking can be induced for example by RGE effects\,\cite{mutau-r:review,nath}.

For hierarchical $N_i$ that we will consider here, $\epsilon = \delta \epsilon$ has to obey the Davidson-Ibarra bound\,\cite{Davidson:2002qv}:
\eq{
	\label{eq:DI_bound}
	|\epsilon| \leq \frac{3}{16\pi} 
	\frac{M_1}{v^2}
	\frac{\Delta m_{hl}^2}
	{\sqrt{\Delta m_{hl}^2+m_l^2} + m_l},
}
where $v = 174$ GeV is the Higgs vacuum expectation value (vev), $m_l\, (m_h)$ is the lightest (heaviest) light neutrino masses and $\Delta m_{hl}^2 \equiv m_h^2 - m_l^2$. In the following, we will fix $\Delta m_{hl}^2$ to the atmospheric mass squared splitting $|\Delta m^2_{\rm atm}| = 2.5 \times 10^{-3}\,{\rm eV}^2$. The upper limit is the largest in the limit $m_l = 0$ for which we have $|\epsilon| \leq 9.9 \times 10^{-5} (M_1/10^{12}\,{\rm GeV})$. Having $m_l$ as large as 0.1 eV, we have instead $|\epsilon| \leq 2.3 \times 10^{-5} (M_1/10^{12}\,{\rm GeV})$.

It is possible to divide the one-flavor regime of temperatures $T\gtrsim 10^{12}\,\unit{GeV}$, roughly into two ranges.
The BEs within these ranges are different but the quantitative consequences are minor.
If leptogenesis happens at $T\gtrsim2\times10^{12}\,{\rm GeV}$ where the EW sphaleron interactions are out of equilibrium\,\cite{Moore:2000mx,Bento:2003jv}, we have 
\begin{eqnarray}
\frac{dY_{L}}{dz} & = & \delta\epsilon\,D\left(\frac{Y_{N}}{Y_{N}^{{\rm eq}}}-1\right)
-\frac{3}{10}D\frac{1}{Y^{{\rm eq}}}Y_{L}\,,
\label{eq:BE1fla_1}
\end{eqnarray}
where $L$ is the total lepton number. 
If leptogenesis happens in this regime, lepton number is projected in $\Delta \equiv B-L = \sum_\alpha \Delta_\alpha$ once EW sphaleron interactions get into equilibrium at $T \lesssim 2\times 10^{12}$ GeV and we have $Y_\Delta = - Y_L$.

If leptogensis happens at $4\times10^{11}\,{\rm GeV}\lesssim T\lesssim2\times10^{12}\,{\rm GeV}$
where the EW sphaleron interactions are already in equilibrium\,\cite{Moore:2000mx,Bento:2003jv}, we should track the total abundance in $Y_\Delta$ by
\begin{eqnarray}
\frac{dY_{\Delta}}{dz} & = & - \delta\epsilon\,D\left(\frac{Y_{N}}{Y_{N}^{{\rm eq}}}-1\right)
- \frac{1}{5}D\frac{1}{Y^{{\rm eq}}}Y_{\Delta} .
\label{eq:BE1fla_2}
\end{eqnarray}

Notice that eq.~\eqref{eq:BE1fla_1} and eq.~\eqref{eq:BE1fla_2} differ by a small numerical factor in the washout term (second terms on the right-hand side). 
In either cases, the final baryon asymmetry $Y_B$ is related to $Y_\Delta$ through eq.~\eqref{eq:YBYDelta}.

\begin{figure}
\begin{centering}
\includegraphics[scale=0.5]{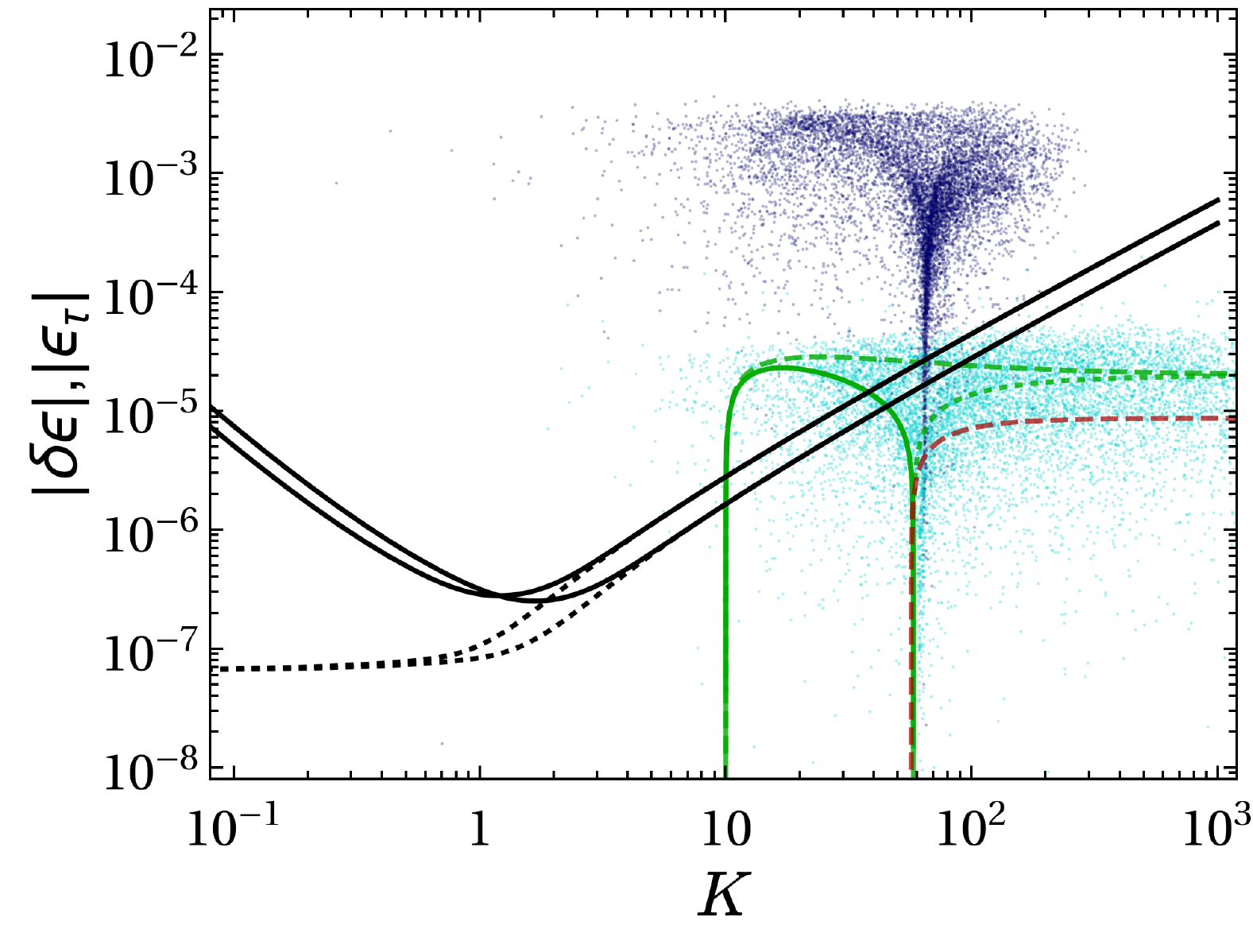}
\par\end{centering}
\caption{\label{fig.1.f}
Required $\left|\delta\epsilon\right|$ (black) to explain the observed baryon
asymmetry in the one flavor regime and possible $|\eps_\tau|$ (colored) with $\cpmutau$ symmetry, both as functions of $K$.
Solid and dashed black lines refer to zero initial $N_{1}$ abundance and thermal initial $N_{1}$ abundance, respectively.
The colored lines indicate the possible values for $|\eps_\tau|$ for hierarchical $N_i$ masses with $M_1=10^{12}\,\unit{GeV}$ and decoupled $N_3$ for the cases of Normal Ordering, NO-(00)/(31) (solid green), NO-(12)/(23) (dashed green),  NO-(13)/(22) (dotted green), and of Inverted Ordering, IO-(11)/(22) (dashed red).
The nomenclature is from ref.\,\cite{cp.mutau} and refer to the CP parities.
The case IO-(12)/(21) is very close to the dashed red line. The omitted line for the case IO-(00)/(33) is very narrow and only covers the bottom part of the red dashed line below $6\times 10^{-8}$. Rescaling $M_1$ up will scale the lines proportionally up.
The scattered points show random points (for all cases with $\cpmutau$ and random lightest neutrino mass) without the assumption of hierarchical $N_i$ for $M_1=10^{12}\,\unit{GeV}$ (cyan) and $M_1=10^{14}\,\unit{GeV}$ (dark blue).
3$\sigma$ ranges from Ref.\,\cite{nufit} were used for the neutrino mixing parameters $\theta_{12},\theta_{13}$ not fixed by $\cpmutau$ and neutrino mass squared differences.
We include the one-loop corrections as discussed in the Sec.\,\ref{sec:large.breaking}.
}
\end{figure}

Using the analytical approximate solution for eq.~\eqref{eq:BE1fla_1} or \eqref{eq:BE1fla_2} shown in Appendix \ref{app:approx_sol}, we obtain the necessary $|\delta\eps|$ to achieve the experimental baryon asymmetry shown in Fig.\,\ref{fig.1.f} (black lines) as a function of the sole relevant variable
\eq{
K \equiv \frac{\Gamma_N}{\cH(z=1)} \equiv \frac{\tm}{m_*}
\,,
}
with $m_*\approx 1\,\unit{meV}$ and
\eq{
\tm =\frac{(\lambda\lambda^\dag)_{11}v^2}{M_1}.
}
In terms of $K$, eq.~\eqref{eq:decay} can be written as
\eq{
\label{eq:DK}
D = K Y_{N}^{\rm eq}(0) z^3 {\cal K}_1(z),
}
with $Y_{N_i}^{\rm eq}(0) = \frac{45}{\pi^4 g_\star}$.
The two solid (dashed) black lines are respectively the solutions of eq.~\eqref{eq:BE1fla_1} and eq.~\eqref{eq:BE1fla_2} assuming zero (thermal) initial $N_1$  abundance. In the $K > 1$ regime, the upper [lower] black line corresponds to the solutions of eq.~\eqref{eq:BE1fla_1} [eq.~\eqref{eq:BE1fla_2}].

To compare the amount of necessary breaking in $\delta\eps$ with the flavored $\eps_\tau$, 
we also show in Fig.\,\ref{fig.1.f} the possible values of $|\eps_\tau|$ in any model of type I seesaw with $\cpmutau$ for different cases. These include the different CP parities that are possible since Majorana phases are trivial\,\cite{cp.mutau}.
In green and red lines we show $|\eps_\tau|$ for the different cases restricted to hierarchical $N_i$ and $N_3$ decoupled case, with $M_1=10^{12}\,\unit{GeV}$, in the low end of the one-flavor regime. 
For this case, we can see that leptogenesis cannot be successful solely with a small breaking of $\cpmutau$.
Masses for $M_1$ of the order $10^{14}\,\unit{GeV}$ are required 
to generate a sufficiently large $|\eps_\tau|$, allowing an $|\delta\eps|$ of 10\% that is still sufficiently large.

Qualitatively, the same conclusion holds even if we allow mild hierarchies for $N_i$ and include the effects of $N_3$ in the loop.
This can be seen from the scatter points in cyan and dark blue that are randomly generated for $M_1=10^{12}\,\unit{GeV}$ and $M_1=10^{14}\,\unit{GeV}$, respectively, 
allowing the mild hierarchy $9M_1\le 3 M_2 \le M_3$.
Only for the region $1\lesssim K\lesssim 10$ and $|\eps_\tau|\gtrsim 10^{-5}$, which contain very few cyan points, can leptogenesis be successful.
We use the Casas-Ibarra parametrization\,\cite{Casas:2001sr} satisfying $\cpmutau$ and exclude large Yukawa couplings $\lambda$ by requiring that $|\lambda_{i\alpha}|\le 1$ so that scatterings which violate lepton number by 2 units can be neglected during leptogenesis.\footnote{%
These scatterings are proportional to $|\lambda_{i\alpha}|^2 |\lambda_{i\beta}|^2$ and will be important when $|\lambda_{i\alpha}| \gtrsim 1$. In this case, the asymmetry generated will generally be too suppressed for successful leptogenesis.}
We use the restriction $|\xi_i|<3$ for parameters appearing in the Casas-Ibarra parametrization. The details of the parametrization are explained in Appendix \ref{app:CI}.

\subsection{Three-flavor regime}
\label{sec:3f:break}

Here we consider the scenario where $M_1 < T_R \lesssim 10^9$ GeV such that $N_1$ leptogenesis occurs in the three-flavor regime. We continue to assume $N_1$ dominated leptogenesis as discussed in the beginning of Sec. \ref{sec:mutau_breaking}.

First let us consider the scenario $P_{\mu}=P_{\tau}$ while $\cpmutau$ is broken in the CP parameters with $\epsilon_{e}=\delta\epsilon_{e}$
and $\epsilon_{\mu}+\epsilon_{\tau}=\delta\epsilon$. Again, we quantify the small breaking by $|\delta\epsilon_e|,|\delta \epsilon| \ll |\epsilon_\tau|$. In this case, we only have to solve the BE for $\vec Y_+$ from eq.~\eqref{eq:BE_matrix_form_p2} with $\delta = \mathbf{0}$:
\eq{
\label{eq:3Feps}
\frac{d \vec{Y}_{+}}{dz} 
= - X_+ \vec{\epsilon}\, D \left(\frac{Y_{N}}{Y_{N}^{{\rm eq}}}-1\right)
+\frac{1}{2Y^{{\rm eq}}}D
\Sigma F \vec{Y}_+ , 
}
where
\begin{eqnarray}
\label{delta.epsilon}
X_+ \vec\epsilon &=& 
(\delta\epsilon_e,\delta\epsilon/2,\delta\epsilon/2)^T, \\
\Sigma &=&
{\rm diag}(P_e,P_\tau,P_\tau).
\end{eqnarray}

In the following, we will consider the following two possibilities: 
\begin{eqnarray}
\mbox{Aligned}: \delta \epsilon_e & = & \delta \epsilon, \label{eq:aligned} \\
\mbox{Purely flavored}: \delta \epsilon_e & = & -\delta \epsilon. \label{eq:pureflavor}
\end{eqnarray}
For the first possibility, $\epsilon = \delta\epsilon_e + \delta\epsilon = 2 \delta\epsilon$ should respect the Davidson-Ibarra bound in eq.~\eqref{eq:DI_bound}. The largest upper bound is obtained for $m_l = 0$ for which we have
$|\epsilon| = |2 \delta \epsilon| \leq 9.9 \times 10^{-8} (M_1/10^9\,{\rm GeV})$. As we will see shortly, leptogenesis can barely be successful in this case. The second possibility is a purely flavored scenario where the total CP asymmetry is zero, $\epsilon = 0$, which trivially satisfies the Davidson-Ibarra bound \eqref{eq:DI_bound}. In this case, there is in principle no bound on $\delta\epsilon$. 

\begin{figure}
	\begin{centering}
		\includegraphics[scale=0.5]{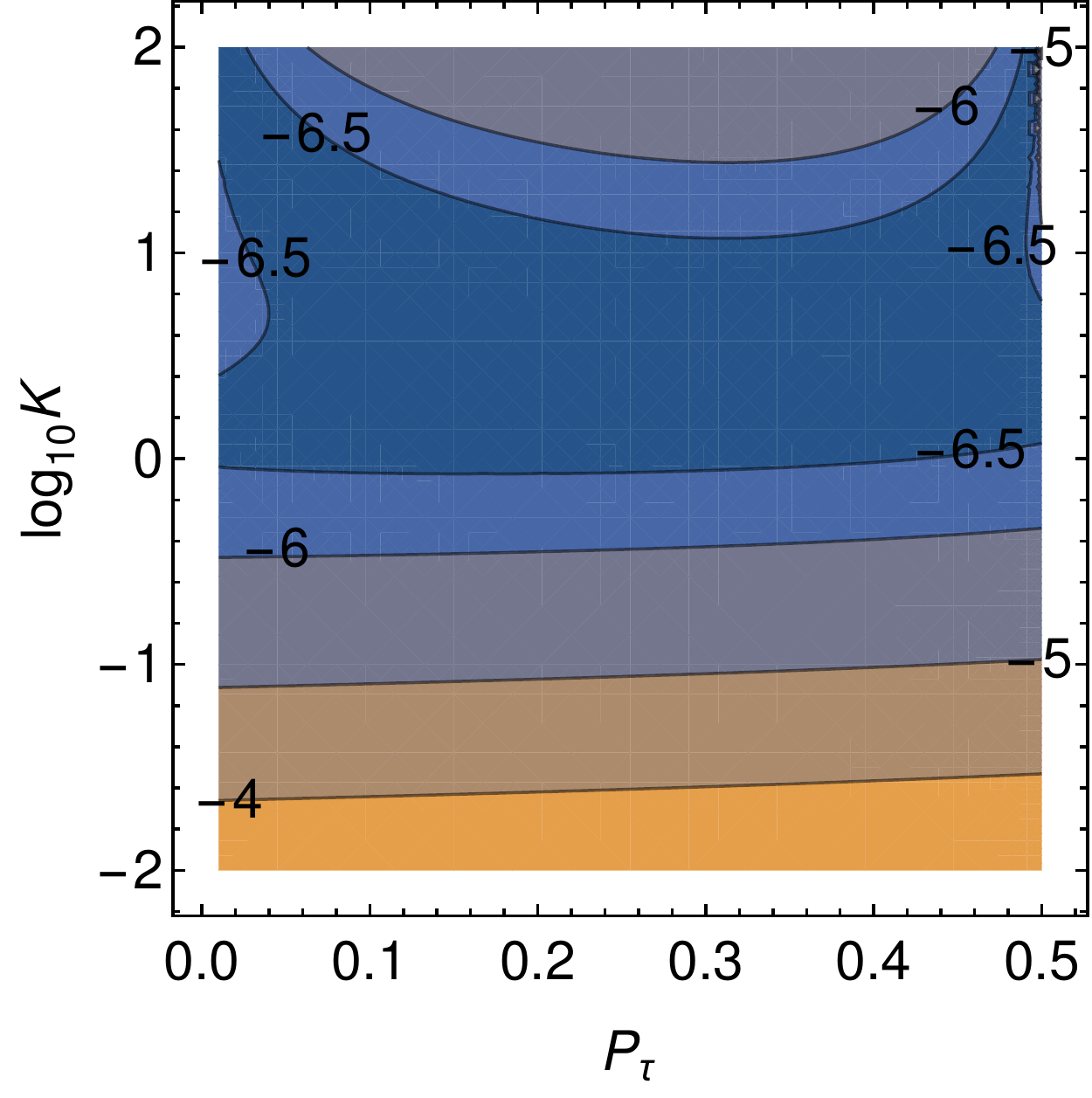}\hspace{0.8cm}\includegraphics[scale=0.5]{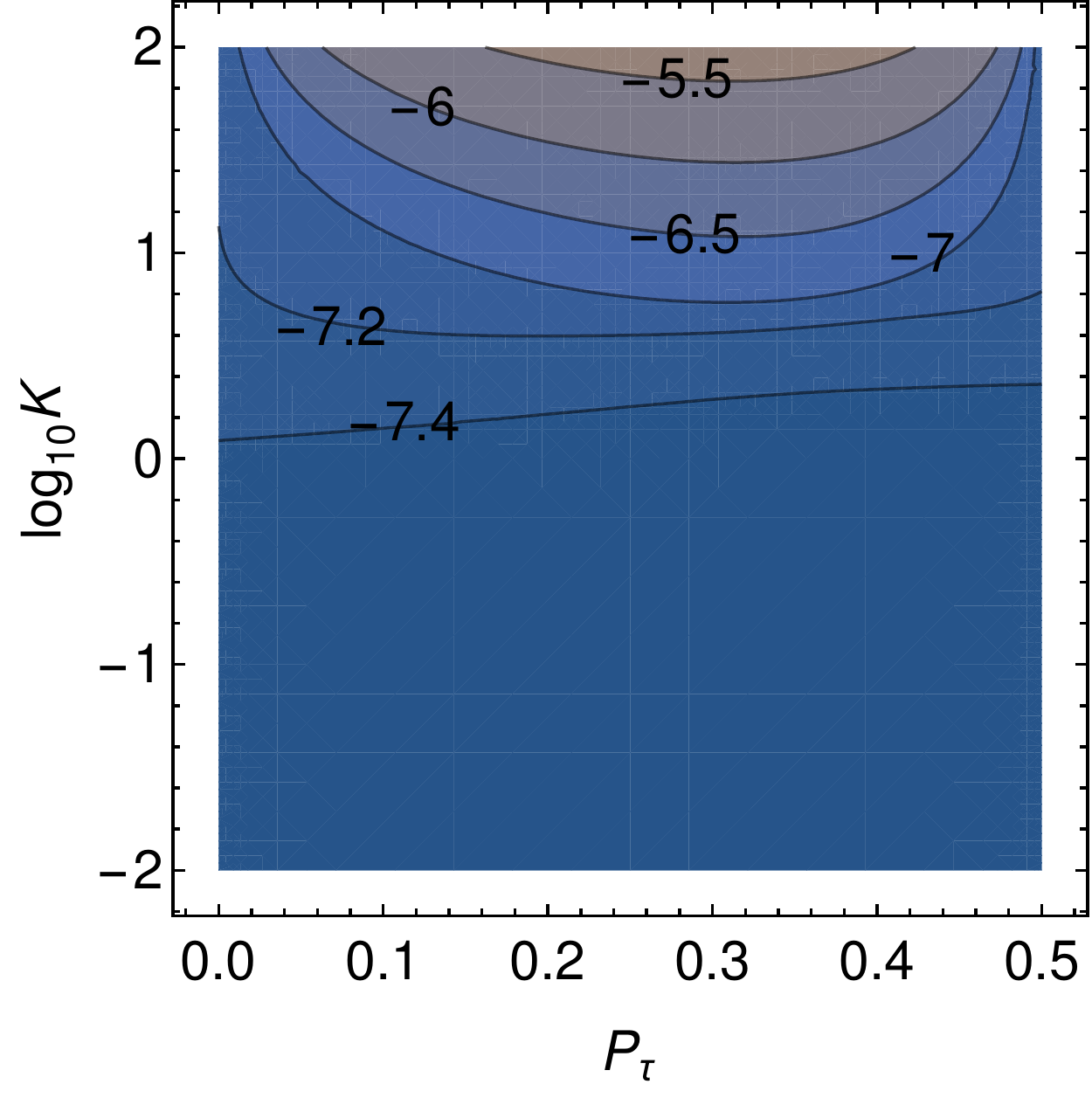}
		\par\end{centering}
	\caption{\label{fig:PP1}
		Required $\log_{10}\left|\delta\epsilon\right|$ for the aligned case $\delta\epsilon_e = \delta\epsilon$ for zero (left) and thermal (right) initial $N_1$ abundance to explain the observed baryon
		asymmetry.
	}
\end{figure}

\begin{figure}
	\begin{centering}
		\includegraphics[scale=0.5]{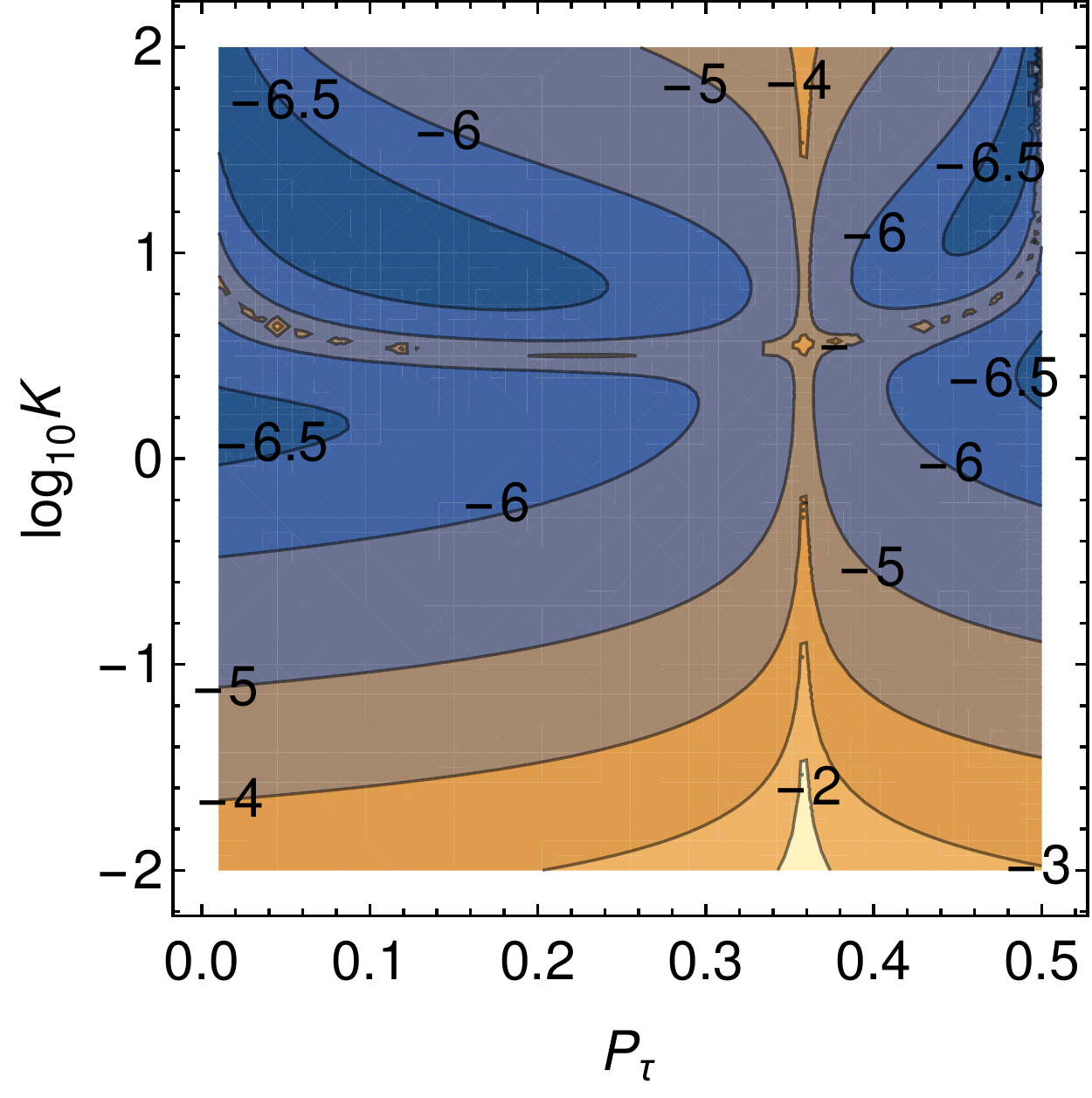}\hspace{0.8cm}
		\includegraphics[scale=0.5]{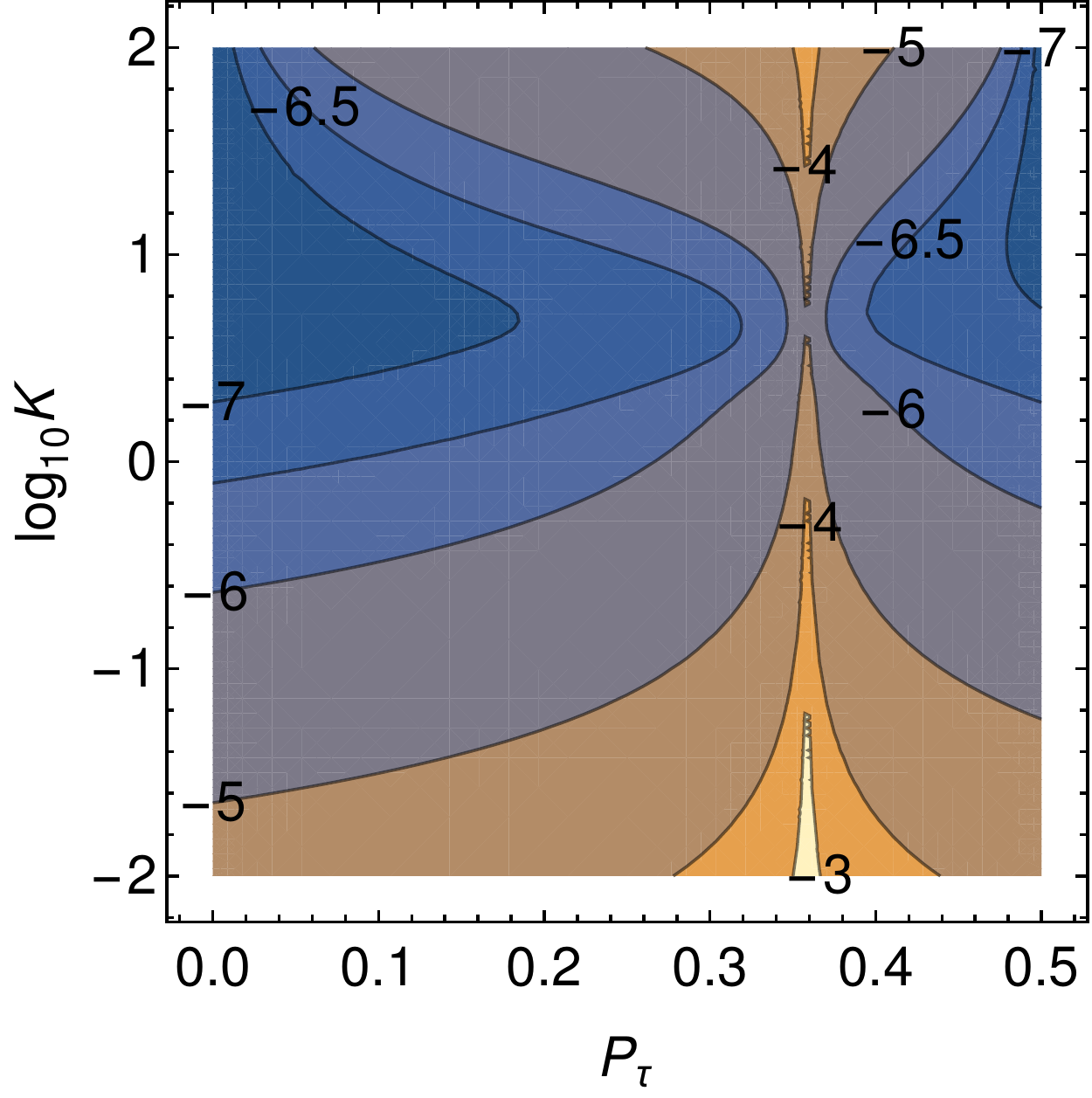}
		\par\end{centering}
	\caption{\label{fig:PP2}
		Required $\log_{10}\left|\delta\epsilon\right|$ for the purely flavored case $\delta\epsilon_e = -\delta\epsilon$ for zero (left) and thermal (right) initial $N_1$ abundance to explain the observed baryon
		asymmetry.
	}
\end{figure}

Applying the solution in eq.~\eqref{eq:sym_result}, 
the required $|\delta\epsilon|$ for the aligned and purely flavored case are plotted respectively in Fig.~\ref{fig:PP1} and Fig.~\ref{fig:PP2} on the plane of $K-P_\tau$. These are the relevant parameters in eq.~\eqref{eq:3Feps} because $D$ is proportional to $K$ as in eq.~\eqref{eq:DK} and $P_e=1-2P_\tau$. 

For the aligned case in Fig.~\ref{fig:PP1}, we observe the following features. For zero initial $N_1$ abundance (left plot), in the weak washout regime $K < 1$, the final $B-L$ asymmetry is suppressed by $K^2$ while in the strong washout regime $K > 1$, it is suppressed by $1/K$. For thermal initial $N_1$ abundance (right plot), in the weak washout regime $K < 1$, the final $B-L$ asymmetry saturate to maximal value while in the strong washout regime $K > 1$, it is also suppressed by $1/K$. 
Since $|\delta \epsilon| \gtrsim 10^{-7}$ is required, leptogenesis can barely be successful due to the Davidson-Ibarra bound discussed below eq.~\eqref{eq:pureflavor}. 

For the purely flavored case in Fig.~\ref{fig:PP2}, we observe that the final $B-L$ asymmetry is suppressed in the weak washout regime $K<1$ for both zero (left plot) and thermal (right plot) initial $N_1$ abundance. This is a specific feature of purely flavored leptogenesis since in the absence of washout, leptogenesis fails because the total CP parameter is null.
Since $|\delta\epsilon|$ is not subjected to the Davidson-Ibarra bound, this scenario can be successful in a larger parameter space.

Next, we will analyze the scenario where $\cpmutau$ is broken only in the flavor projectors $P_{\mu}=P_{\tau}\left(1 - \chi \right)$ with $0 < \chi <1$ while we keep
$\epsilon_{e}=0$ and $\epsilon_{\mu}+\epsilon_{\tau}=0$.\footnote{One can also parametrize $P_{\tau}=P_{\mu}\left(1 - \chi \right)$. In this case, it will correspond to a change of overall sign in the CP parameter $X_- \vec \epsilon \to -X_- \vec \epsilon$.} In this case, the small breaking is quantified by $\chi \ll 1$. From eqs.~\eqref{eq:BE_matrix_form_p2} and \eqref{eq:BE_matrix_form_m2} with $X_+ \vec\epsilon = 0$, we have 
\begin{eqnarray}
\frac{d \vec{Y}_{+}}{dz} 
& = & \frac{1}{2Y^{{\rm eq}}} D
\left( \Sigma F \vec{Y}_+ 
+ \delta F \vec{Y}_- \right), \label{eq:BE_matrix_form_p_break_P} \\
\frac{d \vec{Y}_{-}}{dz} 
& = & - X_- \vec{\epsilon} D\left(\frac{Y_{N}}{Y_{N}^{{\rm eq}}}-1\right)
+ \frac{1}{2Y^{{\rm eq}}}D
\left( \delta F \vec{Y}_+ 
+ \Sigma F \vec{Y}_- \right) ,
\label{eq:BE_matrix_form_m_break_P}
\end{eqnarray}
where $X_- \vec\epsilon$ is given in eq.~\eqref{eq:CP_mutau} and
\begin{eqnarray}
\Sigma & = & 
{\rm diag}\left(P_e,P_\tau (1-\chi/2),P_\tau (1-\chi/2)\right),  \\
\delta & = & \frac{P_\tau \chi}{2}
{\rm diag}\left(0,-1,1\right).
\end{eqnarray}
In the above, we have $P_e = 1 - 2P_\tau(1-\chi/2)$.
Although $\vec{Y}_-$ does not contribute to the $B-L$ asymmetry, it feeds into the BE for $\vec{Y}_+$ through the second term in eq.~\eqref{eq:BE_matrix_form_p_break_P}. 

\begin{figure}
	\begin{centering}
		\includegraphics[scale=0.5]{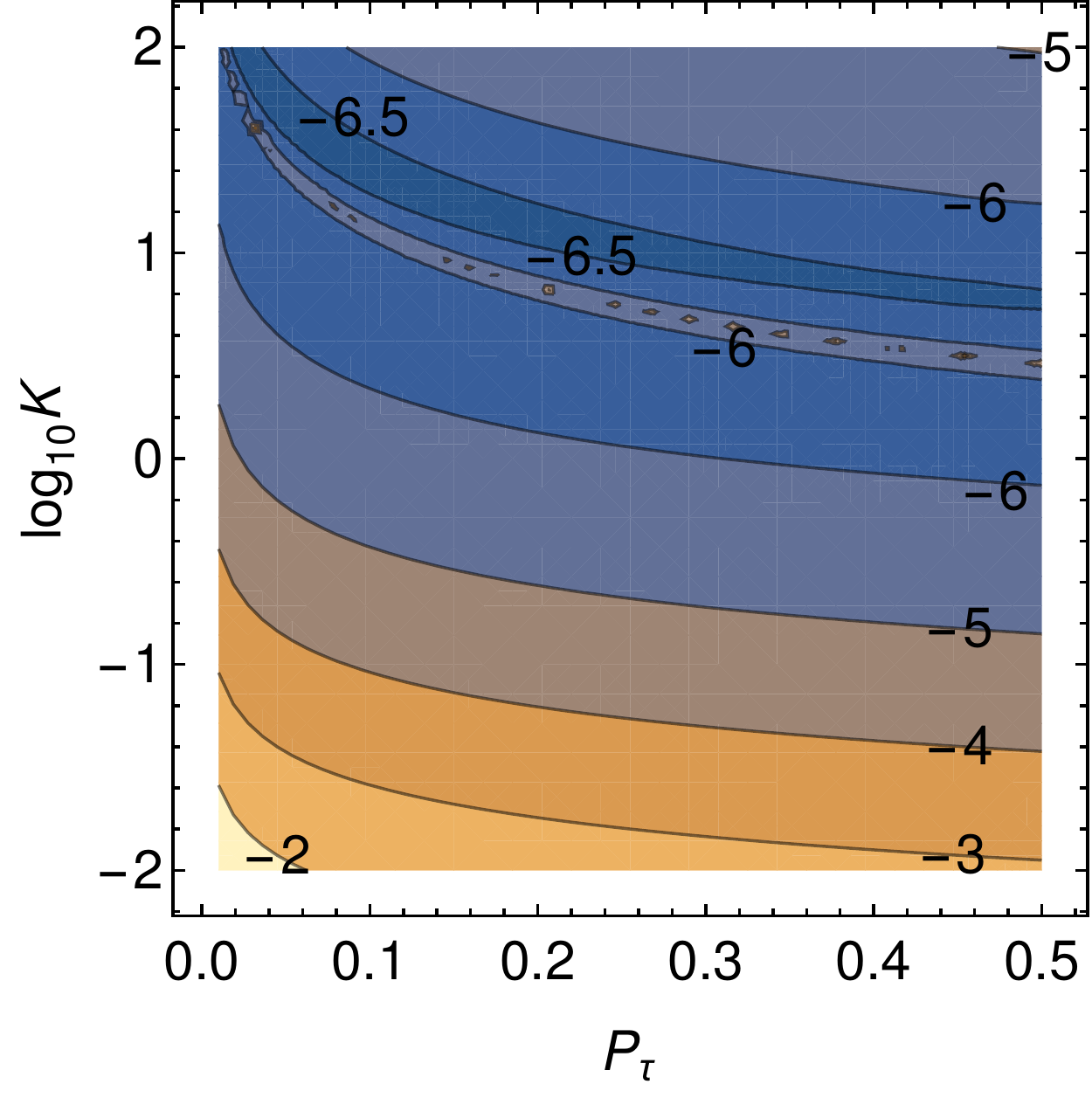}\hspace{0.8cm}\includegraphics[scale=0.5]{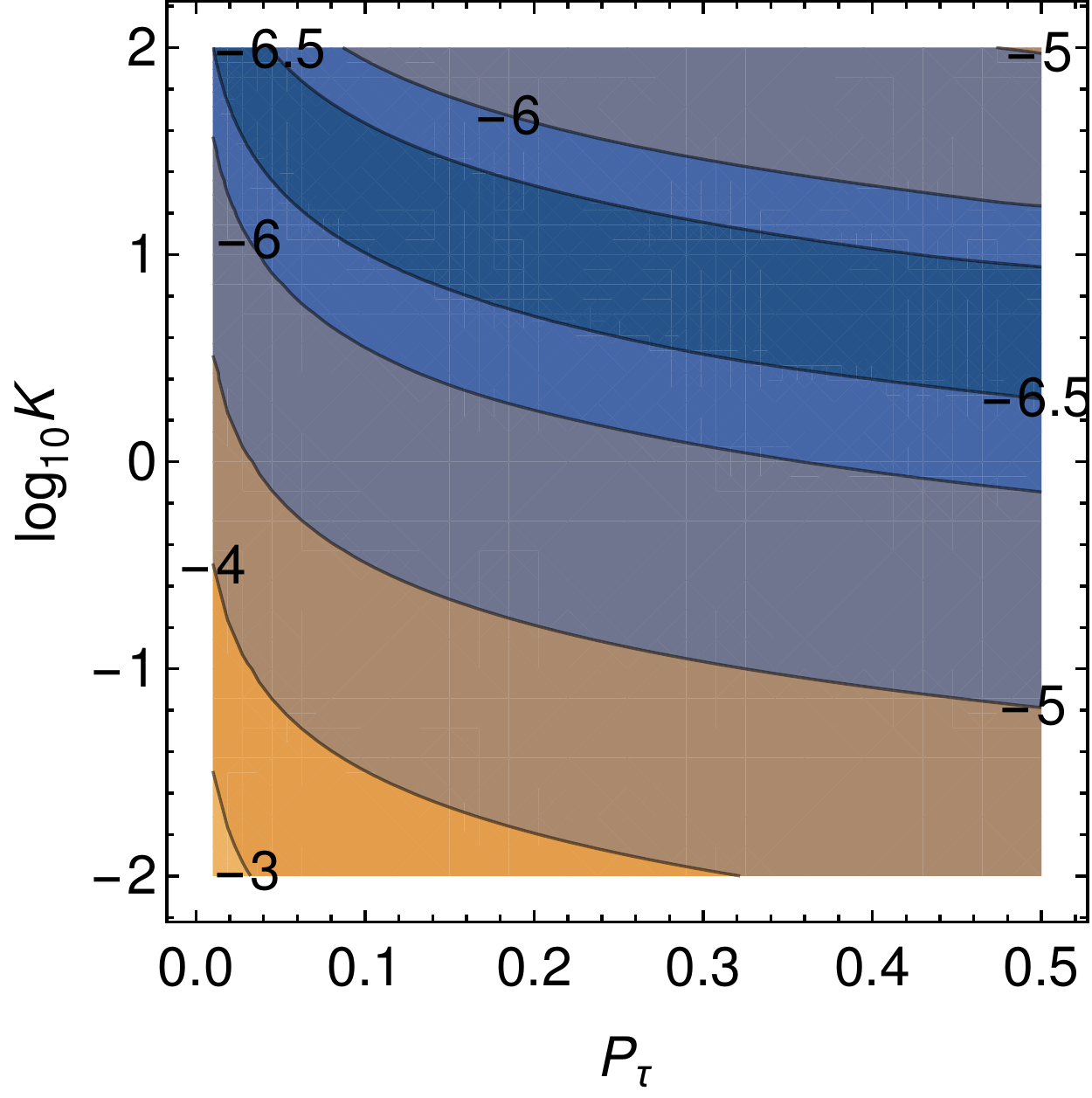}
		\par\end{centering}
	\caption{\label{fig:delta01}
Required $\log_{10}(\chi\left|\epsilon_\tau\right|)$ with small $\cpmutau$ breaking  $\chi = 0.1$ 
to explain the observed baryon asymmetry with zero (left) and thermal (right) initial $N_{1}$ abundance.
	}
\end{figure}

\begin{figure}
	\begin{centering}
		\includegraphics[scale=0.5]{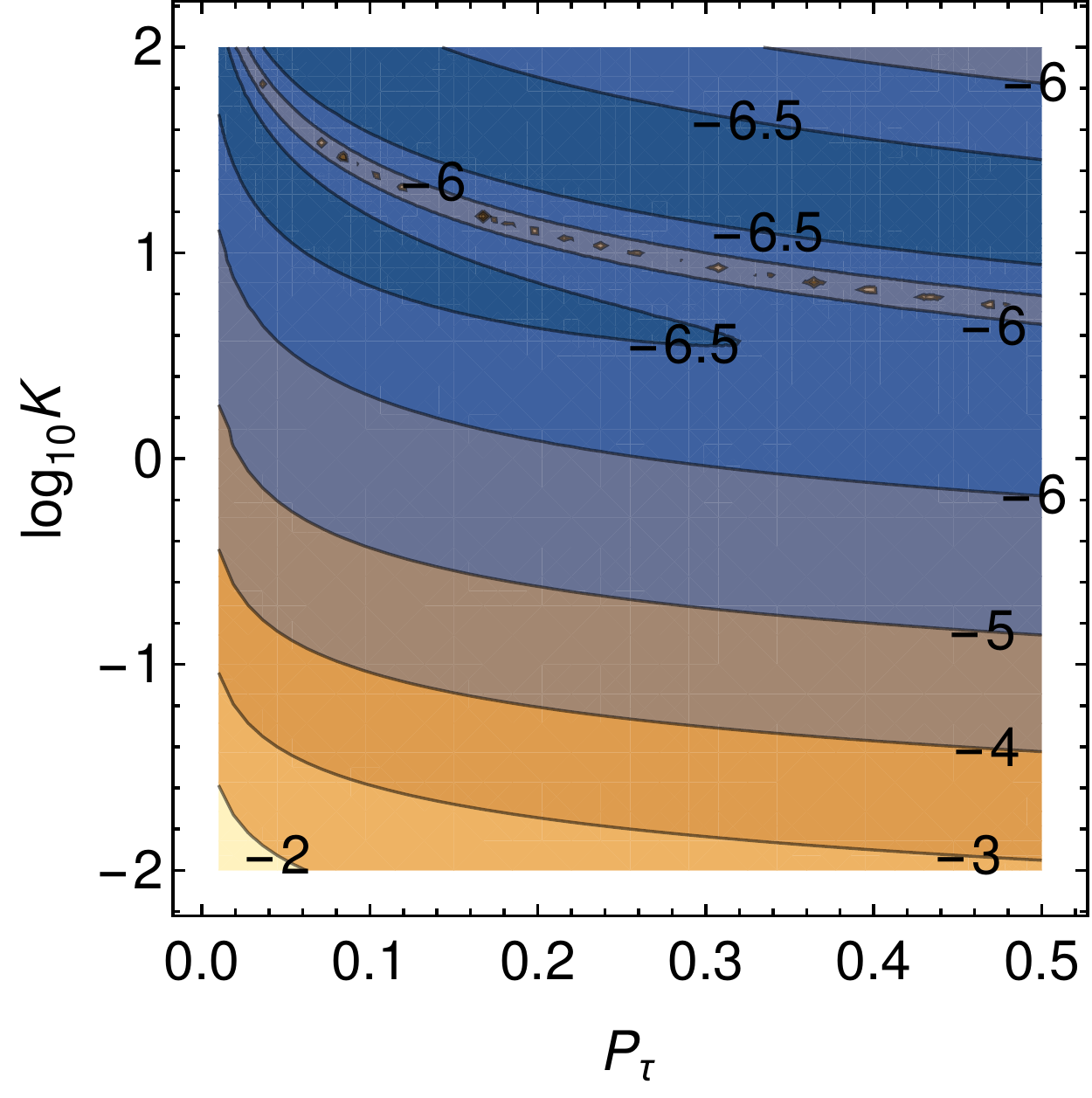}\hspace{0.8cm}\includegraphics[scale=0.5]{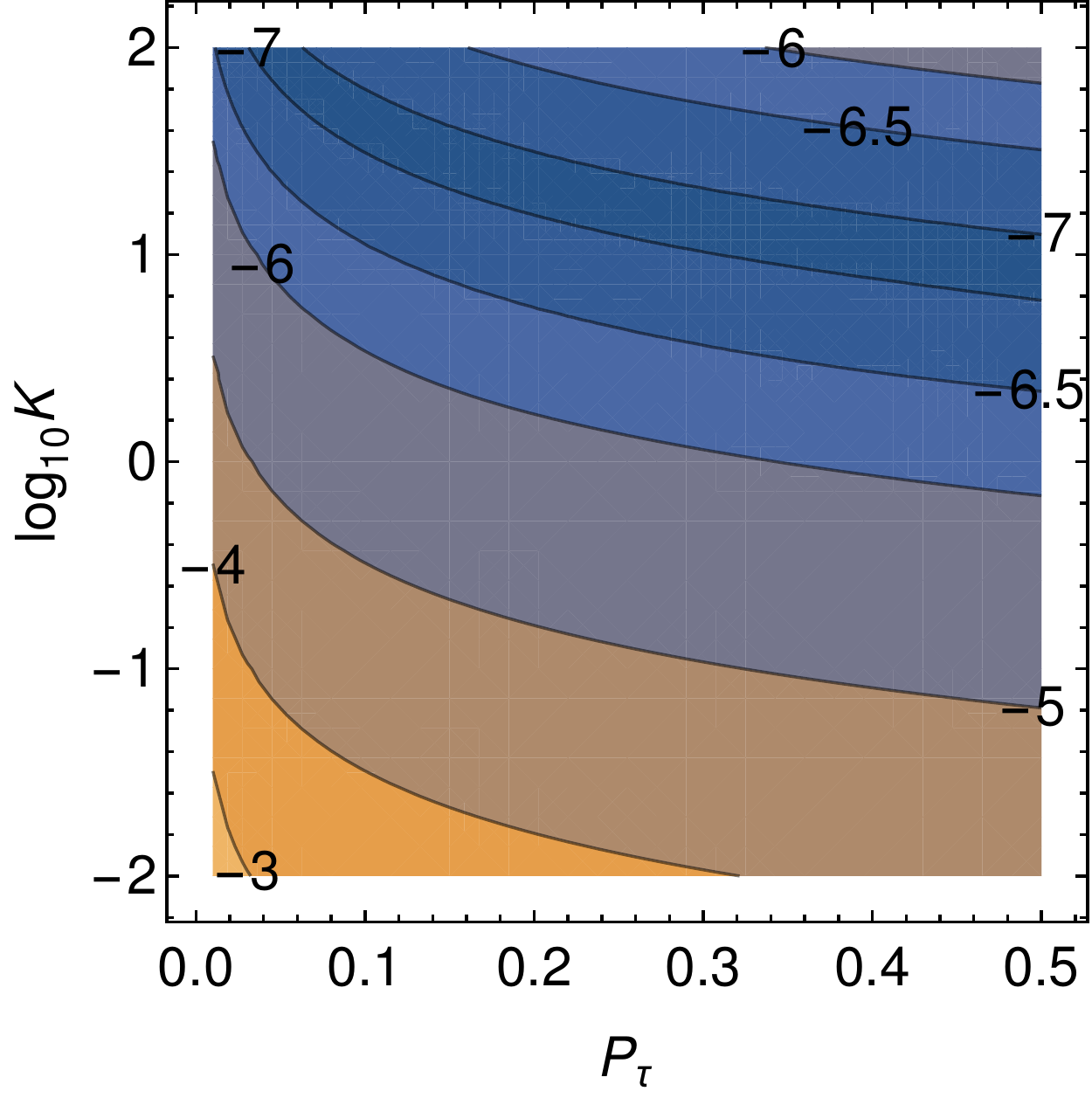}
		\par\end{centering}
	\caption{\label{fig:delta08}
Required $\log_{10}(\chi\left|\epsilon_\tau\right|)$ with large $\cpmutau$ breaking of $\chi = 0.8$ (in log scale) to explain the observed baryon asymmetry with zero (left) and thermal (right) initial $N_{1}$ abundance.
	}
\end{figure}

For $\chi \ll 1$, the final asymmetry is proportional to $\chi \epsilon_\tau$ as shown in eq.~\eqref{eq:perturbed_result}. 
Applying this solution, we plot in Fig.~\ref{fig:delta01} the required $\chi\left|\epsilon_\tau\right|$ for zero (left plot) and thermal (right plot) initial $N_1$ abundance on the $K-P_\tau$ plane for the case $\chi = 0.1$. This case differs very slightly from the case with $\chi < 0.01$ in which the contour of $\chi|\epsilon_\tau| = 10^{-6.5}$ slightly shrinks. For comparison, we also show the case with large $\cpmutau$ breaking in the projectors with $\chi = 0.8$ in Fig.~\ref{fig:delta08}. In this case, eq.~\eqref{eq:perturbed_result} which holds for $\chi \ll 1$, is not a good approximation and we resort to solving eqs.~\eqref{eq:BE_matrix_form_p_break_P} and \eqref{eq:BE_matrix_form_m_break_P} directly.
Notice that leptogenesis is always inefficient in the weak washout regime $K < 1$, i.e., it requires large CP parameters. This is actually a feature of purely flavored leptogenesis as we discussed earlier. Here we can also understand this feature from
eq.~\eqref{eq:BE_matrix_form_p_break_P} 
in which the source term (the second term) 
is proportional to $\frac{1}{2Y^{\rm eq}}D\delta F \vec Y_-$ and reduces as $K$ decreases.

Finally, we can also have $\cpmutau$ breaking in both the CP parameters $\epsilon_e \neq 0, \epsilon_\mu + \epsilon_\tau \neq 0$ and the flavor projectors $P_\mu \neq P_\tau$. Nevertheless, the maximum amount of breaking required will be quantitatively similar and not worth showing. The reader can superimpose Fig.~\ref{fig:PP1} or \ref{fig:PP2} with Fig.~\ref{fig:delta01} or \ref{fig:delta08} to get a sense.

\section{Large $\cpmutau$ breaking}
\label{sec:large.breaking}

Here we quantify large $\cpmutau$ breaking in the CP parameters by
\eq{
	\label{eq:large_break}
		|\delta\eps| \gtrsim |\eps_\tau|\,.
}
For small breaking as in eq.~\eqref{eq:small_break}, 
we have seen in Sec.\,\ref{sec:1f:break} that, for $M_1\approx 10^{12}\,\unit{GeV}$ in the lower end of the one-flavor regime, leptogenesis was barely possible only in the uncommon region $|\eps_\tau|\gtrsim 10^{-5}$ and $1\lesssim K\lesssim 10$; cf.\,Fig.\,\ref{fig.1.f}.
If we allow a large breaking in the form of large $\delta\eps$ in eq.~\eqref{total.eps:break}, leptogenesis is possible in the one-flavor regime for typical parameters as it approaches the generic case.

\begin{figure}
\begin{centering}
\includegraphics[scale=0.45]{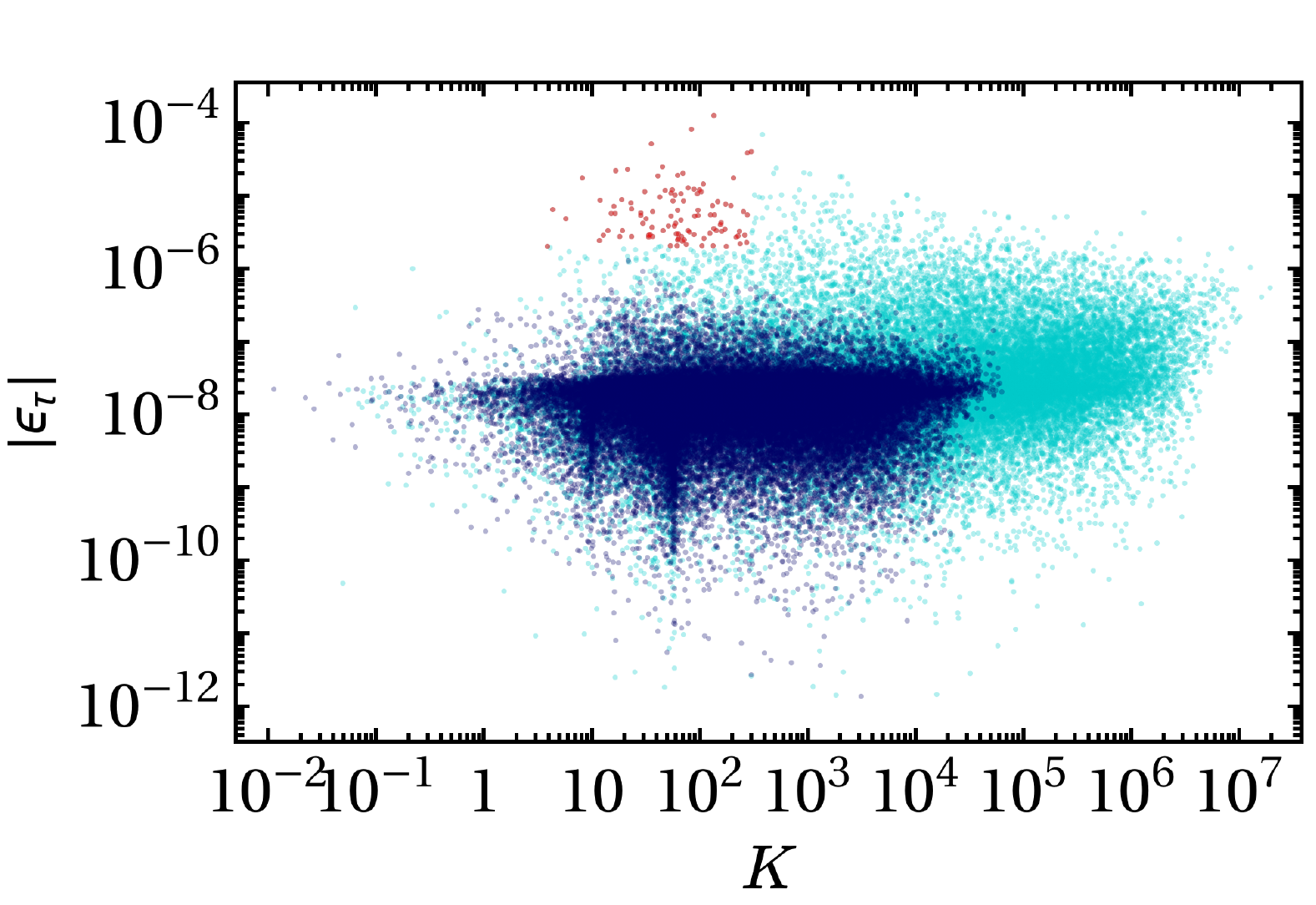}\hspace{0.8cm}
\includegraphics[scale=0.45]{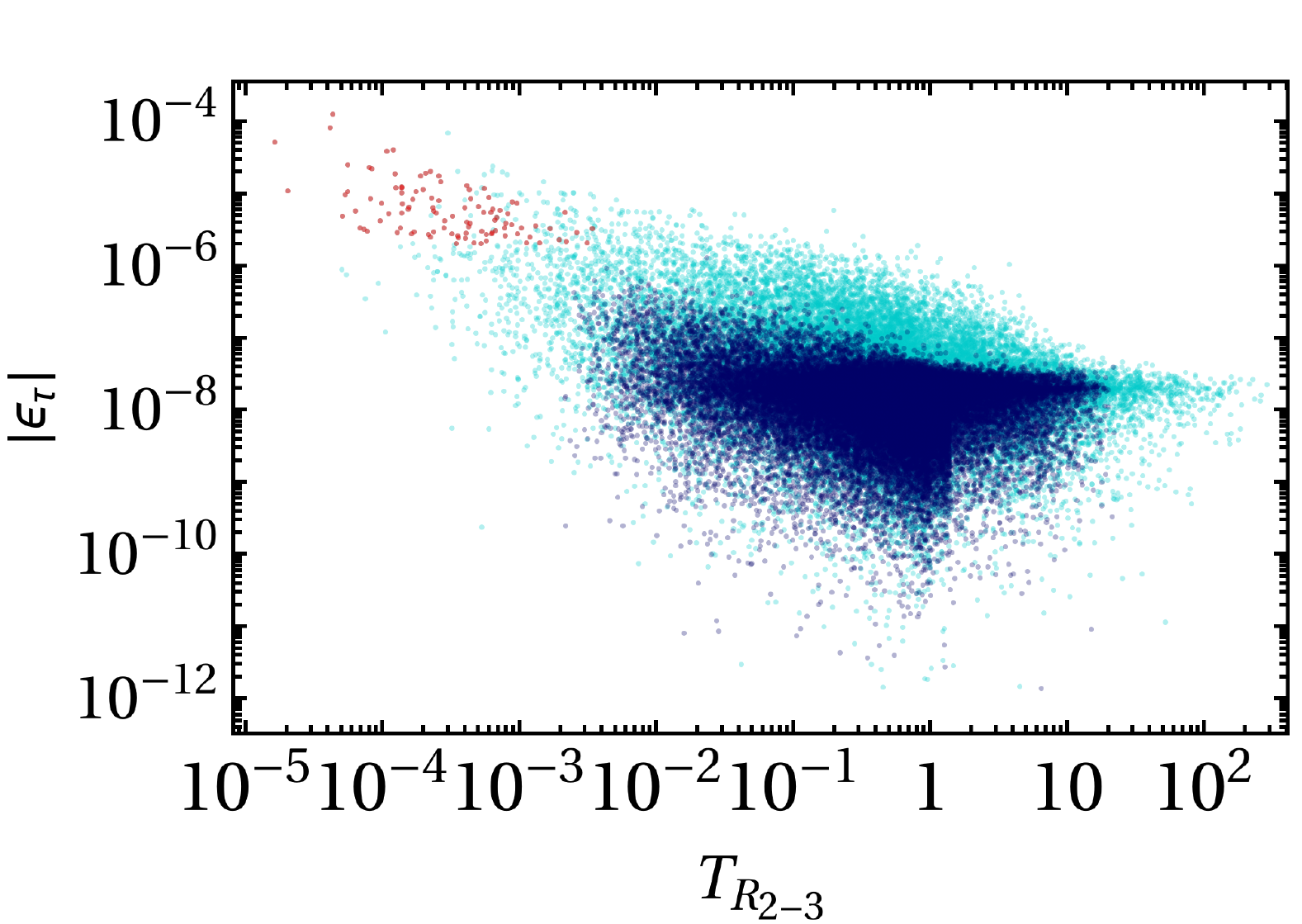}
\par\end{centering}
\caption{\label{fig.3.f}
Scatter plots of $|\eps_\tau|$ against $K$ (left) and the ratio defined in \eqref{tuning:R2-3} (right) for the $\cpmutau$ symmetric model assuming $M_1=10^9\,\unit{GeV}$. Light blue and dark blue indicate different allowed ranges for parameters in the Casas-Ibarra parametrization.
See text for details and also Appendix \ref{app:CI}.
}
\end{figure}

In an analogous manner, in the three-flavor regime analyzed in Sec.\,\ref{sec:3f:break}, a large breaking in the CP asymmetries of at least $|\delta\eps_e|\sim |\delta\eps|\sim 10^{-6}$ (cf. Figs.\,\ref{fig:PP1} and \ref{fig:PP2}) in eq.~\eqref{delta.epsilon} or for a small breaking in the flavor projectors, $(P_\tau-P_\mu)/2\sim 0.1$, a CP asymmetry of at least $|\eps_\tau|\sim 10^{-5}$ was required.
These contrast with the typical value of $|\eps_\tau|\sim 10^{-8}$ shown in the scatter plots of 
Fig.\,\ref{fig.3.f} that show the possible ranges for the $\cpmutau$ symmetric model.\,\footnote{We only show NO but IO is similar.} 
The dark (light) blue points assume $|\xi_i|\le 3$  ($|\xi_i|\le 6$) for parameters $\xi_i$ that appears in the orthogonal matrix $R$ of the Casas-Ibarra parametrization; see explicit parametrization in Appendix \ref{app:CI}.
The red points are a small subset of the light blue points with relatively large $|\eps_\tau|$ and moderate $K$, as can be seen in Fig.\,\ref{fig.3.f} (left).
In contrast,
Fig.\,\ref{fig.3.f} (right) shows $|\eps_\tau|$ against the ratio
\eq{
\label{tuning:R2-3}
T_{R_{2-3}}\equiv 
\frac{\sqrt{\sum_i(|R_{2i}|-|R_{3i}|)^2}}{\sqrt{\sum_i|R_{3i}|^2}}.
}
When this ratio is small, the second and third rows of $R$ are very similar in magnitude and this feature may be regarded as indication of fine-tuning or additional (approximate) symmetry.
Also, when this ratio is small, the first row of $R$ is small in magnitude and this explains how the entries of $R$ may be large but $K$ is kept moderate .
We can see that the red points are concentrated where the ratio is small which shows that fine-tuning is necessary to generate large $\eps_\tau$ but not too large $K$.
We can then conclude that some fine-tuning (or additional symmetry) is required to achieve successful leptogenesis with small $\cpmutau$ breaking.

We can compare the tuning quantified in eq.~\eqref{tuning:R2-3} with other measures.
Considering the type I seesaw formula in the basis $\nu_L\nu_L$, 
\eq{
\label{Mnu:SS}
M_\nu = -m_D^\tp \hat{M}_R^{-1} m_D\,,
}
where $m_D=v\lambda$ and $\hat{M}_R$ is the right-handed neutrino mass matrix which is real, positive and diagonal. The seesaw naturally explains the light neutrino mass of order 0.1\,\unit{eV} for $\lambda\sim 1$ if $\hat{M}_R\sim 10^{14}\,\unit{GeV}$.
Lighter right-handed neutrinos would require smaller $\lambda$ unless some cancellation\,\cite{kersten.smirnov} between $m_D$ and $\hat M_R^{-1}$ takes place in the seesaw formula in eq.~\eqref{Mnu:SS}.
The degree of cancellation can be roughly quantified by 
\eq{
\label{tuning:SS}
T_{SS}\equiv \frac{v^2\|\lambda\|^2\norm{\hat M_R^{-1}}}{\|M_\nu\|}\,,
}
where the double bars denote the matrix norm
\eq{
\|A\|\equiv \sqrt{\operatorname{tr}[AA^\dag]}=\sqrt{\sum_{ij}|A_{ij}|^2}\,.
}
The larger the measure of eq.~\eqref{tuning:SS} the larger the degree of cancellation.
A similar measure is given by some norm of the matrix $R$ in the Casas-Ibarra parametrization\,\cite{aristizabal}.

We will also see that to get large $|\eps_\tau|$, it is essential that we also consider the one-loop contribution to the neutrino mass matrix 
coming from Higgs and $Z$ in the loop:
\eq{
\label{Mnu:1-loop}
M_\nu^{1{-}l}=
m_D^\tp \hat{M}_R^{-1}C_{\rm eff}(\hat{M}_R)m_D\,,
}
where $C_{\rm eff}(\hat{M}_R)$ is a function on the diagonal entries for which\,\cite{Mnu:1-loop}
\eq{
C_{\rm eff}(M)=\frac{M^2}{32\pi^2 v^2}\left(
\frac{\ln(M^2/m_H^2)}{M^2/m_H^2-1}+3\frac{\ln(M^2/m_Z^2)}{M^2/m_Z^2-1}
\right)\,;
}
$m_H$ and $m_Z$ are the masses of the Higgs and the $Z$.
The total neutrino mass matrix is then modified to
\eq{
\label{Mnu:tree+1-l}
M_\nu=M_\nu^{\rm tree}+M_\nu^{1-l}\,.
}
Keeping $\hat{M}_\nu$ as the physical neutrino masses, one can include the one-loop contribution in the Casas-Ibarra parametrization by modifying the Yukawa coupling $\lambda$ to an effective $\lambda^{\rm eff}$\,\cite{casas-ibarra:1-loop}.
See Appendix \ref{app:CI} for the expression.
To quantify the relative contribution from the one-loop contribution, we further use \,\cite{pascoli.turner}
\eq{
\label{tuning:1-loop}
T_{loop}\equiv \frac{\sum_i\operatorname{SVD}(M_\nu^{1-l})_i}{\sum_i\operatorname{SVD}(M_\nu)_i}
\,,
}
where SVD denotes the singular values of the matrix.
The larger this measure, the larger is the one-loop contribution, and if it is much larger than unity, it means that the tree and one-loop contributions are large but they largely cancel each other in \eqref{Mnu:tree+1-l}.

\begin{figure}
\begin{centering}
\includegraphics[scale=0.45]{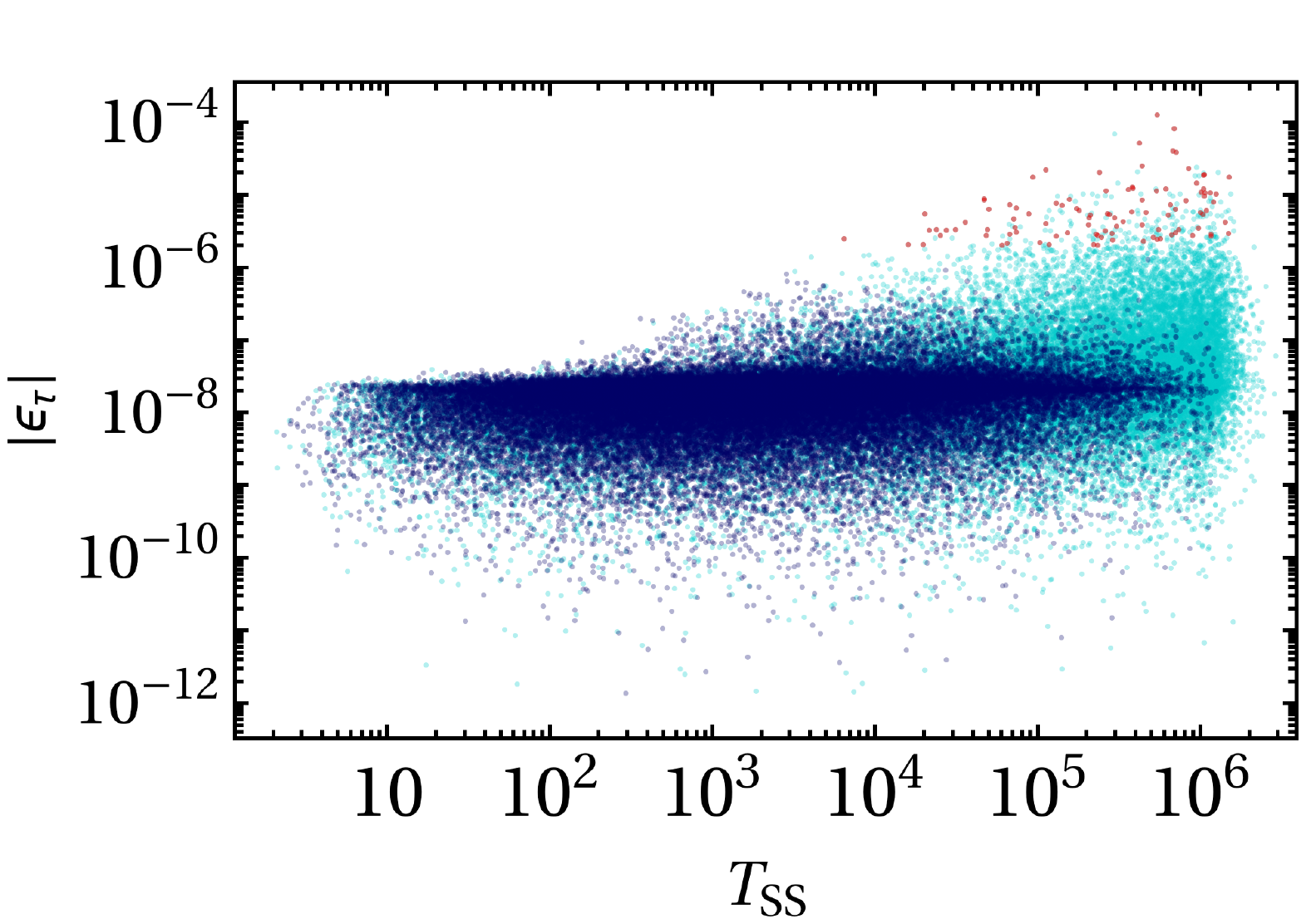}\hspace{0.8cm}
\includegraphics[scale=0.45]{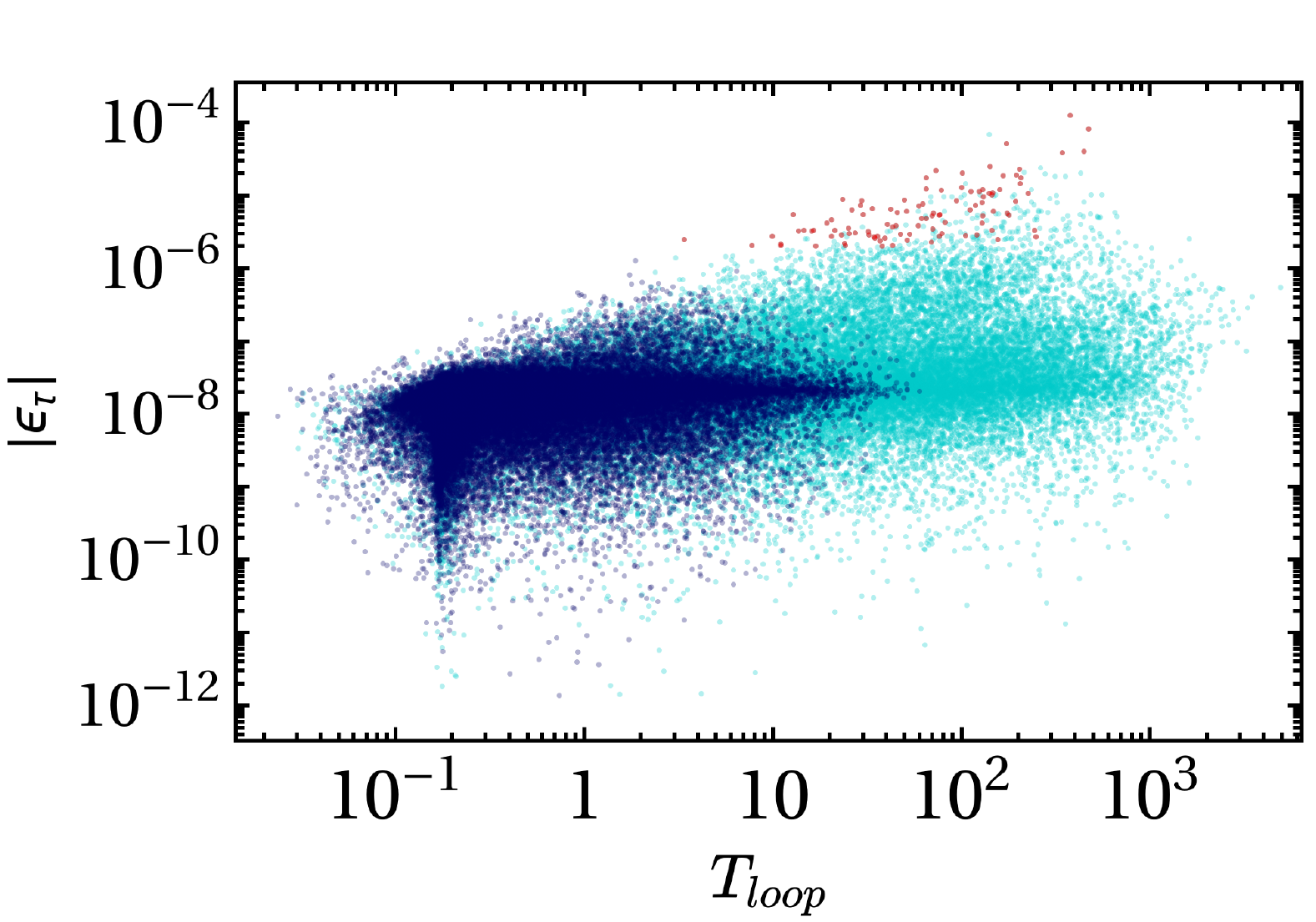}
\par\end{centering}
\caption{\label{fig:tuning}
Scatter plots of $|\eps_\tau|$ against the fine-tuning in the seesaw measured by $T_{SS}$ in \eqref{tuning:SS} (left) and the relative one-loop contribution in $T_{loop}$ in \eqref{tuning:1-loop} (right) for the $\cpmutau$ symmetric model assuming $M_1=10^9\,\unit{GeV}$. Light blue and dark blue indicate different allowed ranges for parameters in the Casas-Ibarra paremetrization.
See text for details and also Appendix \ref{app:CI}.
}
\end{figure}

We illustrate the different fine-tunings needed in Fig.\,\ref{fig:tuning}. The left plot shows how $|\eps_\tau|$ varies with the seesaw cancellation measure \eqref{tuning:SS}, using the effective $\lambda^{\rm eff}$ that already takes the one-loop contribution into account.
We can see that large $|\eps_\tau|\gtrsim 10^{-6}$ (large $|\lambda|$), which includes our red points, is only possible if some cancellation takes place in the seesaw. 
Since a delicate cancellation between $\lambda^{\rm eff}$ and $M_R^{-1}$ is necessary for large $|\eps_\tau|$, the same cancellation is unlikely to happen between $\lambda$ and $M_R^{-1}$ just for the tree-level contribution.
This means that large cancellations between tree and one-loop contributions take place.
We show this in Fig.\,\ref{fig:tuning} (right) where we plot $|\eps_\tau|$ against the measure of eq.~\eqref{tuning:1-loop}.
We also see that $|\eps_\tau|\gtrsim 10^{-6}$ is only possible for large $T_{loop}$.
We note that for leptogenesis with (approximate) $\cpmutau$ a small value for the measure $T_{R_{2-3}}$ \eqref{tuning:R2-3} is a better indicator for successful leptogenesis than $T_{SS}$ or $T_{loop}$ because it allows large $|\eps_\tau|$ with moderate $K$.
Instead, large $T_{SS}$ or $T_{loop}$ tend to lead to large $K$ as well.

To conclude this section, successful leptogenesis in the three-flavor regime with approximate $\cpmutau$ symmetry is only barely possible in the upper end of mass value $M_1 \sim 10^{9}$ GeV if we allow a large degree of cancellation in the neutrino mass matrix in the seesaw and consequently large cancellation between tree and one-loop contributions.
The necessary fine-tuning\footnote{%
In principle one may seek a symmetry justification for the fine-tuning but we do not consider this case.
}
can be alleviated 
by allowing large breaking of $\cpmutau$ in the relevant parameters of leptogenesis, i.e., large difference between $P_\mu$ and $P_\tau$ or/and large deviation from the $\mu\tau$ odd property in eq.~\eqref{eps} for the CP asymmetries.

\section{Examples of large $\cpmutau$ breaking only on {\protect\large$\eps_\alpha$}}
\label{mutau-U:models}

Here we show some examples where large deviations from the $\cpmutau$ symmetric case are possible in the CP asymmetries $\eps_\alpha$ but not on the flavor projectors $P_\alpha$.
These models realize the scenario described in \eqref{eq:3Feps} 
where the flavor projectors still respect \eqref{Pmutau} but the CP asymmetries will not be $\mu\tau$ odd as in \eqref{eps} but exhibit a large $\mu\tau$ even part \eqref{delta.epsilon}.
These examples are based on the generalization of $\cpmutau$ called $\mu\tau$-U symmetry in Ref.\,\cite{king.nishi} where PMNS matrix exhibit
\eq{
\label{mutau-U}
\theta_{23}=45^\circ\,,\quad \delta=\pm 90^\circ,
}
but Majorana phases are non-trivial.
Another characterization is that the PMNS matrix obeys $|(U_\nu)_{i\mu}|=|(U_\nu)_{i\tau}|$, $i=1,2,3$. 
These features may be achieved accidentally by symmetry\,\cite{real.cpmutau}.
But here we need full models with the type I seesaw embedded so that consequences on leptogenesis can be analyzed.

The first example is the Littlest mu-tau seesaw model (LSS$\mu\tau$) of Ref.\,\cite{king.zhou}.
The model is supersymmetric with only two right-handed neutrinos 
and requires a lot of auxiliary fields but at the leptogenesis scale it can be simply described by the simple Dirac mass matrix
\eq{
\bar{m}_D=a
\mtrx{
	0 & \frac{1}{\sqrt{2}} & \frac{1}{\sqrt{2}}\cr
	\frac{1}{\sqrt{11}}&\frac{3}{\sqrt{11}}&\frac{1}{\sqrt{11}}
	}\,,
}
and the heavy singlet Majorana mass matrix
\eq{
\bar{M}_R=M_0\mtrx{1 & \cr & 2\omega},
}
where $\omega=e^{i2\pi/3}$.
The low energy neutrino mass matrix is
\eq{
M_\nu=-\bar{m}_D^\tp \bar{M}_R^{-1}\bar{m}_D
= m_s\mtrx{1&3&1\cr 3&9+11\omega & 3+11\omega \cr 1 & 3+11\omega &1+11\omega}\,,
}
with $m_s=-a^2/(22M_0\omega)$.
This model, with only \textit{one parameter} at low energy, accommodates the low energy neutrino parameters within 3$\sigma$ when RGE corrections are taken into account\,\cite{king.zhou}.
Without corrections, we indeed have $\theta_{23}=45^\circ$ and $\delta=-\pi/2$.
There is also one nontrivial Majorana phase $K_\nu=\diag(1,1,e^{-i\beta/2})$ with $\beta\approx -1.8$.

After rephasing of the singlet neutrino fields, the Dirac mass matrix in the mass basis of $N_R$ is
\eq{
m_D=a
\mtrx{
	0 & \frac{1}{\sqrt{2}} & \frac{1}{\sqrt{2}}\cr
	\frac{\omega}{\sqrt{11}}&\frac{3\omega}{\sqrt{11}}&\frac{\omega}{\sqrt{11}}
	}\,.
}
The phase $\omega$ will induce the necessary CP violation.
It is clear from the first row that $P_{e}=0$ and $P_\mu=P_\tau=1/2$, which respect the $\cpmutau$ property (for $N_1$ but not for $N_2$).
The CP parameters defined in eq.~\eqref{eq:CPfla} will, however, deviate from the $\cpmutau$ symmetric case:
\eq{
	\vec{\eps}=\frac{\sqrt{3}M_0 |m_s|}{4\pi v^2}g(4)\times(0,3,1),
	\label{eq:CP_ex1}
}
where $g(x)$ is given in eq.~\eqref{eq:1loop_fun}.
Using $M_0=10^9\,\unit{GeV}$ and $|m_s|=2.66\,\unit{meV}$, we have
\eq{
\vec{\eps}=-10^{-8}\times(0,3.27,1.09),
}
with $\eps=\sum_\alpha\eps_\alpha=-4.36\times 10^{-8}$ and $K= 54.83$.
Although these CP parameters are too small for the three-flavor regime, it certainly may generate the right baryon asymmetry in the one- or two-flavor regimes.

The second example was proposed in Ref.\,\cite{king.nishi} and is also based on constrained sequential dominance (see Ref.\,\cite{king:review} for a review), a supersymmetric model where the Dirac mass matrix has special forms due to vev alignments from nonabelian flavor symmetry.
We assume the charged lepton mass matrix squared $M_l^\dag M_l$ is invariant by
\eq{
\label{T:Z3}
T=\mtrx{0&0&1\cr 1&0&0 \cr 0&1&0}
\,,
}
a generator of groups such as $A_4$ or $S_4$, so that in the flavor basis the charge leptons contribute the mixing matrix $U_l^\dag$ having the form
\eq{
\label{Uw}
U_\om\equiv\frac{1}{\sqrt{3}}\mtrx{1&1&1\cr 1&\om&\om^2\cr 1&\om^2&\om}\,.
}

There are three right-handed neutrinos $N_{iR}$ with diagonal mass matrix
\eq{
\bar M_R=\mtrx{M_1 &&\cr &M_2e^{i\eta_1} & \cr && M_3e^{i\eta_2}}\,,
}
where we allow for phases and the masses are not necessarily ordered from lighter to heavier.
Special flavon vevs give the Dirac mass matrix the special structure
\eq{
\bar{m}_D=\mtrx{au_1^\tp\cr bu_2\tp \cr cu_3^\tp}=\mtrx{a&&\cr &b&\cr &&c}\mtrx{0&1&1\cr 1&-1&1\cr2&-1&1}\,.
}
Note that $u_1\perp u_2,u_3$ and such a structure is easily obtained from vev alignments in indirect models\,\cite{indirect:cp} from, e.g., the $A_4$ symmetry in the CSD framework\,\cite{Bjorkeroth:2014vha} in the real triplet basis of $A_4$.
The parameters $a,b,c$ are chosen to be real by appropriate rephasing.
In the mass basis for the charged leptons and right-handed neutrinos, this matrix becomes
\eq{
m_D=\mtrx{a&&\cr &b\,e^{-i\eta_1/2}&\cr &&c\,e^{-i\eta_2/2}}\mtrx{0&1&1\cr 1&-1&1\cr2&-1&1}
U_\omega^\dag
\,.
}
One can easily check that $P_{i\mu}=P_{i\tau}$.
Moreover, $m_D$ obeys 
\eq{
\label{mD:conj}
m_D^*=\mtrx{1&&\cr&e^{i\eta_1}&\cr&&e^{i\eta_2}}m_DX\,,
}
which generalizes the $\cpmutau$ symmetric case\,\cite{cp.mutau}.

In the symmetry basis of \eqref{T:Z3}, the neutrino mass matrix is given by
\eq{
\label{bar-M:indirect}
M_\nu=m_a u_1u_1^\tp +m_bu_2u_2^\tp + m_cu_3u_3^\tp\,,
}
where
\eq{
m_a=-\frac{a^2}{M_1}\,,\quad
m_b=-e^{-i\eta_1}\frac{b^2}{M_2}\,,\quad
m_c=-e^{-i\eta_2}\frac{c^2}{M_3}\,.
}

If $\eta_2-\eta_1=0,\pi$, one can check that the real and imaginary parts of $M_\nu$ commute so that it is diagonalized by a real orthogonal matrix, leading to a PMNS matrix obeying $\mu\tau$-U\,\cite{king.nishi}.
Additionally, if 
\eq{
m_a\approx 6\,\text{meV}\,,\quad
m_b\approx e^{i\eta}\times 34\,\text{meV}\,,\quad
m_c\approx e^{i\eta}\times(-11)\,\text{meV}\,,
}
we obtain NO spectrum with observables within 3$\sigma$ and lightest mass $m_1\approx 
12\,\text{meV}$.
This example was shown in Ref.\,\cite{king.nishi}.
The phase $\eta$ can take any value and a nonzero value leads to nontrivial Majorana phases without disrupting the predictions of $\mu\tau$-U mixing.
In addition, the mixing obeys the TM1 form\,\cite{TM1}.

However, the additional Majorana phase does not induce a deviation of $\eps_{\alpha}$ from the $\cpmutau$ symmetric form in eq.~\eqref{eq:CP_mutau}.
The reason is the following: because $u_1$ is orthogonal to $u_2,u_3$, the loop contribution from $N_2,N_3$ to the CP asymmetries of $N_1$ vanish.\,\footnote{%
This is analogous to form dominance models corresponding to all $u_i$ real and orthogonal\,\cite{no.lepto.form}.
}
So $N_2$ or $N_3$ needs to be the lightest one and only the interference between $N_2$ and $N_3$ leads to nonzero CP asymmetries.
Due to the property \eqref{mD:conj}, one can show that $\eps_\mu=-\eps_\tau$ from an analogous proof shown in Ref.\,\cite{cp.mutau}.

Defining $\eta_{21} \equiv \eta_2 - \eta_1$ and assuming $M_2 \ll M_3, M_1$, the CP parameters defined in eq.~\eqref{eq:CPfla} for $i=2$ are given by
\eq{
	\vec{\eps} = -\frac{M_2 |m_c|}{12\pi v^2}
	\left( 1 - \frac{2}{3}\frac{M_2}{M_3} \right)
	(2 \sin\eta_{21},\sqrt{3}\cos\eta_{21}+5 \sin\eta_{21},-\sqrt{3}\cos\eta_{21}+5 \sin\eta_{21}).\label{eq:CP_ex2}
}
To induce a deviation from the $\mu\tau$ odd form in \eqref{eq:CP_mutau}, we need $\eta_2-\eta_1\neq 0,\pi$, which also leads to deviation from $\mu\tau$-U in \eqref{mutau-U} but $P_\mu=P_\tau$ is still valid.
For example, we can take $(M_2,M_1,M_3)=10^9(1,10,100)\,\unit{GeV}$ and
\eq{
m_a\approx 8\,\text{meV}\,,\quad
m_b\approx e^{i\pi/3}\times 36.72\,\text{meV}\,,\quad
m_c\approx -e^{i(\pi/3-0.1)}\times(12.87)\,\text{meV}\,,
}
to obtain neutrino oscillation observables within $3\sigma$ and
\eq{
\vec{\eps}\approx 10^{-8}\times (0.23,2.49,-1.37)\,,
}
resulting in $\sum\epsilon_\alpha\approx 1.35\times 10^{-8}$ with $\tilde{m}_\alpha\approx(12.24,49,49)\,\unit{meV}$.
Although not sufficient for successful three-flavor leptogenesis, this is another example of large $\cpmutau$ breaking on $\eps_\alpha$ maintaining $P_\mu=P_\tau$.
In order to maximize the CP parameters in eq.~\eqref{eq:CP_ex2}, one can show that $|m_c| \leq \sqrt{|\Delta m^2_{\rm atm}|}/6$ and hence we have
\eq{
	|\eps_\mu + \eps_\tau| \leq \frac{5M_2 \sqrt{|\Delta m^2_{\rm atm}|}}{36\pi v^2}
	= 2.4 \times 10^{-8}\left(\frac{M_2}{10^9\,{\rm GeV}}\right),
	\label{eq:CPmax_ex2}
}
setting $|\Delta m^2_{\rm atm}| = 2.5 \times 10^{-3}\,{\rm eV}^2$. These CP parameters are clearly too small for the three-flavor regime while it could work naturally for one- or two-flavor regimes.

\section{Conclusions}
\label{sec:concl}

We develop the BEs for leptogenesis in the three-flavor regime in the $\mu\tau$ basis where the description of approximate $\mu\tau$ reflection ($\cpmutau$) symmetric case is particularly simple.
Effectively for leptogenesis, the symmetric case implies equality between the flavor projectors into $\mu$ and $\tau$ flavors and odd behavior of the CP asymmetries under $\mu\tau$ exchange: $\eps_{i\tau}=-\eps_{i\mu}$ and $\eps_{ie}=0$. 
The formalism is useful for all cases where these properties are (approximately) valid and makes use of the fact that the interactions relevant for leptogenesis in the primordial plasma are symmetric by exchange of $\mu\leftrightarrow \tau$.
With the formalism, barring preexisting asymmetry, we confirm previous results that leptogenesis cannot be successful in the three-flavor regime in the exact symmetry limit, even if flavor effects are fully taken into account.

With small breaking of $\cpmutau$, $N_1$ leptogenesis can be barely successful in the three-flavor regime only if we allow large fine-tunings that we quantify using some measures.
The fine-tuning is necessary to push the CP asymmetry $|\eps_\tau|$ to higher values keeping the washout parameter $K$ moderate.
These values can be achieved in our case only if the second and third rows (associated to $N_2,N_3$) of the orthogonal matrix of the Casas-Ibarra parametrization are very similar in magnitude.
Additionally, we confirm that this requires large cancellations in the seesaw formula, calculated using one-loop corrections, between the Dirac mass term and the heavy right-handed neutrino mass matrix.
Barring some protective symmetry, large cancellations between tree and one-loop contribution to the neutrino mass matrix are generically required as well.

With large breaking of $\cpmutau$, the parameter space for leptogenesis in the three-flavor regime widens and less fine-tuning is necessary. 
We end by discussing some examples in which the symmetry is broken only in the CP asymmetries but not in the flavor projectors. These examples can lead to successful leptogenesis in the one- or two-flavor regime but the CP asymmetries turn out to be insufficient for the three-flavor regime.

\acknowledgements

C.C.N.\ acknowledges partial support by Brazilian FAPESP, grant 2014/19164-6, and CNPq, grant 308578/2016-3. C.S.F. acknowledges support by FAPESP, grant 2019/11197-6 and CNPq, grant 301271/2019-4.

\appendix
\section{Flavor coefficients $A$ and $C$} 
\label{app:A_C}

The coefficients which relate particle asymmetries 
to the charges $Y_{\Delta_\alpha}$ can be derived using equilibrium conditions as done in\,\cite{Nardi:2006fx,Abada:2006ea} or more directly from symmetry principle\,\cite{Fong:2015vna}.
In the temperature regime $10^{6}\,{\rm GeV}\lesssim T\lesssim10^{9}\,{\rm GeV}$ where only $u$, $d$, $e$ Yukawa interactions are out of thermal equilibrium, we have
\begin{eqnarray}
A & = & \frac{1}{2148}\left(\begin{array}{ccc}
-6\times151 & 120 & 120\\
75 & -688 & 28\\
75 & 28 & -688
\end{array}\right),\\
C & = & -\frac{1}{716}\left(\begin{array}{ccc}
37 & 52 & 52\end{array}\right).
\end{eqnarray}
In the temperature regime $10^{4}\,{\rm GeV}\lesssim T\lesssim10^{6}\,{\rm GeV}$ where only $e$ Yukawa interaction is out of thermal equilibrium, we have
\begin{eqnarray}
A & = & \frac{1}{2886}\left(\begin{array}{ccc}
-11\times111 & 156 & 156\\
111 & -910 & 52\\
111 & 52 & -910
\end{array}\right),\\
C & = & -\frac{1}{962}\left(\begin{array}{ccc}
37 & 52 & 52\end{array}\right).
\end{eqnarray}
In the temperature regime $10^{2}\,{\rm GeV}\lesssim T\lesssim10^{4}\,{\rm GeV}$ where all Yukawa interactions are in thermal equilibrium, we have
\begin{eqnarray}
A & = & \frac{1}{711}\left(\begin{array}{ccc}
-221 & 16 & 16\\
16 & -221 & 16\\
16 & 16 & -221
\end{array}\right),\\
C & = & -\frac{4}{79}\left(\begin{array}{ccc}
1 & 1 & 1\end{array}\right).
\end{eqnarray}

\section{Analytical approximate solutions in the $\mu\tau$ basis} \label{app:approx_sol_mutau}

In the $\mu\tau$ basis, the general Boltzmann equations for leptogenesis
in 3-flavor regime are given by eqs.~\eqref{eq:BE_matrix_form_p2} and \eqref{eq:BE_matrix_form_m2} in which we rewrite them here for convenience
\begin{eqnarray}
\frac{d\vec{Y}_{\Delta_{+}}}{dz} & = & -\sum_{i}\left[X_{+}\vec{\epsilon}_{i}D_{i}\left(\frac{Y_{N_{i}}}{Y_{N_{i}}^{{\rm eq}}}-1\right)-\frac{1}{2Y^{{\rm eq}}}D_{i}\left(\Sigma_{i}F\vec{Y}_{\Delta_{+}}+\delta_{i}F\vec{Y}_{\Delta_{-}}\right)\right],\\
\frac{d\vec{Y}_{\Delta_{-}}}{dz} & = & -\sum_{i}\left[X_{-}\vec{\epsilon}_{i}D_{i}\left(\frac{Y_{N_{i}}}{Y_{N_{i}}^{{\rm eq}}}-1\right)-\frac{1}{2Y^{{\rm eq}}}D_{i}\left(\delta_{i}F\vec{Y}_{\Delta_{+}}+\Sigma_{i}F\vec{Y}_{\Delta_{-}}\right)\right].
\end{eqnarray}
We can diagonalize the matrix $\Sigma_{i}F$ as follows
\begin{eqnarray}
V_{i}^{-1}\Sigma_{i}FV_{i} & = & {\rm diag}\left(r_{io},r_{i-},r_{i+}\right),
\end{eqnarray}
where the eigenvalues are\footnote{In the following, we have decomposed $F$ into the components of $A$ and $C$. }
\begin{eqnarray}
r_{i\pm} & \equiv & \frac{\left(A_{ee}+C_{e}\right)P_{ie}+\left(A_{\mu\mu}+A_{\mu\tau}+2C_{\mu}\right)P_{i\mu\tau}\pm\sqrt{w_{i}}}{2},\\
r_{io} & \equiv & \left(A_{\mu\mu}-A_{\mu\tau}\right)P_{i\mu\tau},
\end{eqnarray}
with
\begin{eqnarray}
P_{i\mu\tau} & \equiv & \frac{1}{2}\left(P_{i\mu}+P_{i\tau}\right),\\
w_{i} & \equiv & \left[\left(A_{ee}+C_{e}\right)P_{ie}-\left(A_{\mu\mu}+A_{\mu\tau}+2C_{\mu}\right)P_{i\mu\tau}\right]^{2}\nonumber \\
&  & +8\left(A_{e\mu}+C_{e}\right)\left(A_{\mu e}+C_{\mu}\right)P_{ie}P_{i\mu\tau}.
\end{eqnarray}
The matrix $V_{i}$ is given by
\begin{eqnarray}
V_{i} & = & \left(\begin{array}{ccc}
0 & u_{i-} & u_{i+}\\
-1 & 1 & 1\\
1 & 1 & 1
\end{array}\right),
\end{eqnarray}
where
\begin{eqnarray}
u_{i\pm} & \equiv & \frac{\left(A_{ee}+C_{e}\right)P_{ie}-\left(A_{\mu\mu}+A_{\mu\tau}+2C_{\mu}\right)P_{i\mu\tau}\pm\sqrt{w_{i}}}{2\left(A_{\mu e}+C_{\mu}\right)P_{i\mu\tau}}.
\end{eqnarray}
Transforming $\delta_{i}F$ with $V_{i}$, we obtain
\begin{eqnarray}
V_{i}^{-1}\delta_{i}FV_{i} & = & \left(\begin{array}{ccc}
0 & b_{i-} & b_{i+}\\
a_{i-} & 0 & 0\\
a_{i+} & 0 & 0
\end{array}\right),
\end{eqnarray}
where
\begin{eqnarray}
a_{i\pm} & \equiv & \delta P_{i\mu\tau}\left(A_{\mu\mu}-A_{\mu\tau}\right)\left[\pm\frac{\left(A_{ee}+C_{e}\right)P_{ie}-\left(A_{\mu\mu}+A_{\mu\tau}+2C_{\mu}\right)P_{i\mu\tau}}{2\sqrt{w_{i}}}-\frac{1}{2}\right],\\
b_{i\pm} & \equiv & \delta P_{i\mu\tau}\frac{\left[\mp\sqrt{w_{i}}-\left(A_{ee}+C_{e}\right)P_{ie}-\left(A_{\mu\mu}+A_{\mu\tau}+2C_{\mu}\right)P_{i\mu\tau}\right]}{2P_{i\mu\tau}},
\end{eqnarray}
with
\begin{eqnarray}
\delta P_{i\mu\tau} & \equiv & \frac{1}{2}\left(P_{i\mu}-P_{i\tau}\right).
\end{eqnarray}

Assuming that leptogenesis is dominated by decays of particular generation of $N_{i}$, it is useful to transform $\vec{Y}_{\Delta_{\pm}}$ to the following basis\footnote{The inverse of $V_{i}$ is given by
	\begin{eqnarray*}
		V_{i}^{-1} & = & \frac{1}{u_{i+}-u_{i-}}\left(\begin{array}{ccc}
			0 & -\frac{u_{i+}-u_{i-}}{2} & \frac{u_{i+}-u_{i-}}{2}\\
			-1 & \frac{1}{2}u_{i+} & \frac{1}{2}u_{i+}\\
			1 & -\frac{1}{2}u_{i-} & -\frac{1}{2}u_{i-}
		\end{array}\right).
	\end{eqnarray*}
}
\begin{eqnarray}
V_{i}^{-1}\vec{Y}_{\Delta_{+}} & \equiv & \left(\begin{array}{c}
0\\
Y_{i-}\\
Y_{i+}
\end{array}\right),\\
V_{i}^{-1}\vec{Y}_{\Delta_{-}} & \equiv & \left(\begin{array}{c}
Y_{o}\\
0\\
0
\end{array}\right),
\end{eqnarray}
where we have defined
\begin{eqnarray}
Y_{i\pm} & \equiv & \pm\frac{1}{u_{i+}-u_{i-}}\left[Y_{\Delta_{e}}-\frac{1}{2}\left(Y_{\Delta_{\mu}}+Y_{\Delta_{\tau}}\right)u_{i\mp}\right],\\
Y_{o} & \equiv & -\frac{1}{2}\left(Y_{\Delta_{\mu}}-Y_{\Delta_{\tau}}\right).
\end{eqnarray}
Notice that the $Y_{\Delta_{\alpha}}$ asymmetry can be recovered
from the $\mu\tau$ even components in the new basis as follows
\begin{eqnarray}
Y_{\Delta_{e}} & = & u_{i-}Y_{i-}+u_{i+}Y_{i+},\\
Y_{\Delta_{\mu}}+Y_{\Delta_{\tau}} & = & 2\left(Y_{i-}+Y_{i+}\right).
\end{eqnarray}
In this new basis, we have
\begin{eqnarray}
\frac{dY_{i\pm}}{dz} & = & -\left[\epsilon_{i\pm}D_{i}\left(\frac{Y_{N_{i}}}{Y_{N_{i}}^{{\rm eq}}}-1\right)-\frac{1}{2Y^{{\rm eq}}}D_{i}\left(r_{i\pm}Y_{i\pm}+a_{i\pm}Y_{o}\right)\right],\\
\frac{dY_{o}}{dz} & = & -\left[\epsilon_{io}D_{i}\left(\frac{Y_{N_{i}}}{Y_{N_{i}}^{{\rm eq}}}-1\right)-\frac{1}{2Y^{{\rm eq}}}D_{i}\left(b_{i-}Y_{i-}+b_{i+}Y_{i+}+r_{io}Y_{o}\right)\right],
\end{eqnarray}
where we have defined
\begin{eqnarray}
\epsilon_{i\pm} & \equiv & \pm\frac{1}{u_{i+}-u_{i-}}\left[\epsilon_{ie}-\frac{1}{2}\left(\epsilon_{i\mu}+\epsilon_{i\tau}\right)u_{i\mp}\right],\\
\epsilon_{io} & \equiv & -\frac{1}{2}\left(\epsilon_{i\mu}-\epsilon_{i\tau}\right).
\end{eqnarray}
In the case where $P_{i\mu}=P_{i\tau}\implies a_{i\pm},b_{i\pm}=0$,
we only need to solve
\begin{eqnarray}
\frac{dY_{i\pm}}{dz} & = & -\left[\epsilon_{i\pm}D_{i}\left(\frac{Y_{N_{i}}}{Y_{N_{i}}^{{\rm eq}}}-1\right)-\frac{1}{2Y^{{\rm eq}}}D_{i}r_{i\pm}Y_{i\pm}\right].
\end{eqnarray}
The analytical approximate solution in eq.~\eqref{eq:approx_sol} can be
applied directly and we have
\begin{eqnarray}
Y_{i\pm}\left(\infty\right) & = & Y_{i\pm}\left(0\right)e^{\frac{3\pi}{8}Rr_{i\pm}K_{i}}-\epsilon_{i\pm}Y_{N_{i}}^{{\rm eq}}\left(0\right)\eta\left(K_{i},r_{i\pm}\right).
\label{eq:sym_result}
\end{eqnarray}
If $\mathsf{CP}^{\mu\tau}$ is not broken in the CP parameters, $\epsilon_{i\pm}=0$
and the final asymmetry is vanishing in the absence of preexisting
asymmetry $Y_{i\pm}\left(0\right)=0$ in accordance to the result in Sec. \ref{sec:exact_mutau}.

\vspace{1cm}
\subsection{Perturbative diagonalization for $\left|\delta P_{i\mu\tau}\right|\ll1$ \label{app:approx_sol_mutau_pert}}

If $\left|\delta P_{i\mu\tau}\right|\ll1$, we can carry out the perturbative diagonalization and write down the solution at leading order in $\delta P_{i\mu\tau}$. In the basis $\left\{ Y_{i+},Y_{i-},Y_{o}\right\} $, we need to diagonalize $R_{i}=R_{i}^{\left(0\right)}+\delta R_{i}$ where
\begin{eqnarray}
R_{i}^{\left(0\right)} & = & \left(\begin{array}{ccc}
r_{i+} & 0 & 0\\
0 & r_{i-} & 0\\
0 & 0 & r_{io}
\end{array}\right),\\
\delta R_{i} & = & \left(\begin{array}{ccc}
0 & 0 & a_{i+}\\
0 & 0 & a_{i-}\\
b_{i+} & b_{i-} & 0
\end{array}\right).
\end{eqnarray}
While $\delta R_i$ is proportional to $\delta P_{i\mu\tau}$, $R_{i}^{\left(0\right)}$ is already diagonal and therefore the perturbed matrix of eigenvectors for $R$ will be $I+\delta U$. Denoting the perturbed eigenvalues matrix as $\delta r_i={\rm diag}\left(\delta r_{i+},\delta r_{i-},\delta r_{io}\right)$,
we have
\begin{eqnarray}
\left(R_{i}^{\left(0\right)}+\delta R_{i}\right)\left(I+\delta U\right) & = & \left(I+\delta U\right)\left(R_{i}^{\left(0\right)}+\delta r_i\right),\nonumber \\
R_{i}^{\left(0\right)}+\delta R_{i}+R_{i}^{\left(0\right)}\delta U+\delta R_{i}\delta U & = & R_{i}^{\left(0\right)}+\delta r_i+\delta UR_{i}^{\left(0\right)}+\delta U\delta r_i.
\end{eqnarray}
Keeping only the leading terms, we obtain
\begin{eqnarray}
\delta r_i & = & \delta R_{i}+\left[R_{i}^{\left(0\right)},\delta U\right].
\end{eqnarray}
Writing $\left[R_{i}^{\left(0\right)}\delta U\right]_{mn}=r_{m}\left[\delta U\right]_{mn}$
and $\left[\delta UR_{i}^{\left(0\right)}\right]_{mn}=\left[\delta U\right]_{mn}r_{n}$ where we denote $r_1 = r_{i+}$, $r_2 = r_{i-}$ and $r_3 = r_{io}$, we have
\begin{eqnarray}
\left[\delta r_i\right]_{mn} & = & \left[\delta R_{i}\right]_{mn}+\left(r_{m}-r_{n}\right)\left[\delta U\right]_{mn}.
\end{eqnarray}
Setting $m=n$, we have
\begin{eqnarray}
\left[\delta r_{i}\right]_{nn} & = & 0,
\end{eqnarray}
and therefore the eigenvalues are not perturbed at the order $\delta P_{i\mu\tau}$. For $m\neq n$, we obtain
\begin{eqnarray}
0 & = & \left[\delta R_{i}\right]_{mn}+\left(r_{m}-r_{n}\right)\left[\delta U\right]_{mn}\nonumber \\
\implies\left[\delta U\right]_{mn} & = & -\frac{\left[\delta R_{i}\right]_{mn}}{r_{m}-r_{n}}.
\end{eqnarray}
Explicitly, we have
\begin{eqnarray}
\delta U & = & -\left(\begin{array}{ccc}
0 & 0 & \frac{a_{i+}}{r_{i+}-r_{io}}\\
0 & 0 & \frac{a_{i-}}{r_{i-}-r_{io}}\\
-\frac{b_{i+}}{r_{i+}-r_{io}} & -\frac{b_{i-}}{r_{i-}-r_{io}} & 0
\end{array}\right).
\end{eqnarray}

Using the results above, the BEs with perturbative diagonalization up to order $\delta P_{i\mu\tau}$ are given by
\begin{eqnarray}
\frac{d\tilde{Y}_{i\pm}}{dz} & = & -\left[\tilde{\epsilon}_{i\pm}D_{i}\left(\frac{Y_{N_{i}}}{Y_{N_{i}}^{{\rm eq}}}-1\right)-\frac{1}{2Y^{{\rm eq}}}D_{i}r_{i\pm}\tilde{Y}_{i\pm}\right],\\
\frac{d\tilde{Y}_{o}}{dz} & = & -\left[\tilde{\epsilon}_{io}D_{i}\left(\frac{Y_{N_{i}}}{Y_{N_{i}}^{{\rm eq}}}-1\right)-\frac{1}{2Y^{{\rm eq}}}D_{i}r_{io}\tilde{Y}_{o}\right],
\end{eqnarray}
where
\begin{eqnarray}
	\tilde{Y}_{i\pm} & = & Y_{i\pm}+Y_{io}\frac{a_{i\pm}}{r_{i\pm}-r_{io}},\\
	\tilde{Y}_{o} & = & Y_{o}-Y_{i+}\frac{b_{i+}}{r_{i+}-r_{io}}-Y_{i-}\frac{b_{i-}}{r_{i-}-r_{io}},\\
	\tilde{\epsilon}_{i\pm} & = & \epsilon_{i\pm}+\epsilon_{io}\frac{a_{i\pm}}{r_{i\pm}-r_{io}},\\
	\tilde{\epsilon}_{io} & = & \epsilon_{io}-\epsilon_{i+}\frac{b_{i+}}{r_{i+}-r_{io}}-\epsilon_{i-}\frac{b_{i-}}{r_{i-}-r_{io}}.
\end{eqnarray}
The analytical approximate solution in eq.~\eqref{eq:approx_sol} can be
applied directly and we obtain
\begin{eqnarray}
\tilde{Y}_{i\pm}\left(\infty\right) & = & \tilde{Y}_{i\pm}\left(0\right)e^{\frac{3\pi}{8}Rr_{i\pm}K_{1}}-\tilde{\epsilon}_{i\pm}Y_{N_{i}}^{{\rm eq}}\left(0\right)\eta\left(K_{i},r_{i\pm}\right),\\
\tilde{Y}_{o}\left(\infty\right) & = & \tilde{Y}_{o}\left(0\right)e^{\frac{3\pi}{8}Rr_{io}K_{1}}-\tilde{\epsilon}_{io}Y_{N_{i}}^{{\rm eq}}\left(0\right)\eta\left(K_{i},r_{io}\right).
\end{eqnarray}

Let us consider the case of vanishing initial asymmetries $\tilde{Y}_{i\pm}\left(0\right)=\tilde{Y}_{o}\left(0\right)=0$.
Transforming back to the basis of $Y_{i\pm}$, we have
\begin{eqnarray}
Y_{i\pm}\left(\infty\right) & = & -\left(\epsilon_{i\pm}+\epsilon_{io}\frac{a_{i\pm}}{r_{i\pm}-r_{io}}\right)Y_{N_{i}}^{{\rm eq}}\left(0\right)\eta\left(K_{i},r_{i\pm}\right)+\frac{a_{i\pm}}{r_{i\pm}-r_{io}}\epsilon_{io}Y_{N_{i}}^{{\rm eq}}\left(0\right)\eta\left(K_{i},r_{io}\right)\nonumber \\
& = & -\epsilon_{i\pm}Y_{N_{i}}^{{\rm eq}}\left(0\right)\eta\left(K_{i},r_{i\pm}\right)-\frac{a_{i\pm}}{r_{i\pm}-r_{io}}\epsilon_{io}Y_{N_{i}}^{{\rm eq}}\left(0\right)\left[\eta\left(K_{i},r_{i\pm}\right)-\eta\left(K_{i},r_{io}\right)\right]. \label{eq:perturbed_full_result}
\end{eqnarray}
If $\mathsf{CP}^{\mu\tau}$ is not broken in the CP parameters, $\epsilon_{i\pm}=0$ and $\epsilon_{io}=\epsilon_{i\tau}$, and the final asymmetry becomes
\begin{eqnarray}
Y_{i\pm}\left(\infty\right) & = & -\frac{a_{i\pm}}{r_{i\pm}-r_{io}}\epsilon_{i\tau}Y_{N_{i}}^{{\rm eq}}\left(0\right)\left[\eta\left(K_{i},r_{i\pm}\right)-\eta\left(K_{i},r_{io}\right)\right],
\label{eq:perturbed_result}
\end{eqnarray}
which is proportional to $\delta P_{i\mu\tau}\epsilon_{i\tau}$.

\section{Analytical approximate solutions} 
\label{app:approx_sol}

For the BE of the form
\begin{eqnarray}
\frac{dY_{\Delta}}{dz} & = & S\left(z\right)+W\left(z\right)Y_{\Delta},
\end{eqnarray}
the solution is
\begin{eqnarray}
Y_{\Delta}\left(z\right) & = & Y_{\Delta}\left(z_{0}\right)e^{\int_{z_{0}}^{z}dz'W\left(z'\right)}+\int_{z_{0}}^{z}dz'S\left(z'\right)e^{r\int_{z'}^{z}dz''W\left(z''\right)},
\end{eqnarray}
where $z_{0}$ is some initial value of the variable $z$.

For leptogenesis including only decay and inverse decay processes
of $N_{i}$ with mass $M_{i}$, we have
\begin{eqnarray}
S\left(z\right) & = & -\sum_{i}\epsilon_{i}D_{i}\left(\frac{Y_{N_{i}}}{Y_{N_{i}}^{{\rm eq}}}-1\right)=\sum_{i}\epsilon_{i}\frac{dY_{N_{i}}}{dz},\\
W\left(z\right) & = & \frac{1}{2Y^{{\rm eq}}}\sum_{i}r_{i}D_{i}.
\end{eqnarray}
where $\epsilon_{i}$ and $r_{i}$ are parameters independent of $z$,
and $Y^{{\rm eq}}=\frac{15}{8\pi^{2}g_{\star}}$ with $g_{\star}$
the cosmic total relativistic degrees of freedom. Choosing $z\equiv\frac{M_{1}}{T}$,
we have
\begin{eqnarray}
Y_{N_{i}}^{{\rm eq}} & = & \frac{45}{2\pi^{4}g_{\star}}a_{i}^{2}z^{2}{\cal K}_{2}\left(a_{i}z\right),\\
D_{i} & = & Y_{N_{i}}^{{\rm eq}}\frac{\Gamma_{N_{i}}}{{\cal H}z}\frac{{\cal K}_{1}\left(a_{i}z\right)}{{\cal K}_{2}\left(a_{i}z\right)},
\end{eqnarray}
where $a_{i}\equiv\frac{M_{i}}{M_{1}}$, ${\cal K}_{n}\left(x\right)$
is the modified Bessel function of the second kind of order $n$,
$\Gamma_{N_{i}}=\frac{\left(\lambda\lambda^{\dagger}\right)_{ii}M_{i}}{8\pi}$ is the $N_{i}$ total decay width, and ${\cal H}=1.66\sqrt{g_{\star}}\frac{T^{2}}{M_{{\rm Pl}}}$ is the Hubble rate with $M_{{\rm Pl}}=1.22\times10^{19}$ GeV.

Assuming leptogenesis to be dominated by the decays and inverse decays
of only $N_{1}$ and taking the initial temperature to be very large
$z_{0}\to0$, the final asymmetry at $z\to\infty$ can be approximated
by
\begin{eqnarray}
Y_{\Delta}\left(\infty\right) & = & Y_{\Delta}\left(0\right)e^{\frac{3\pi}{8}Rr_{1}K_{1}}-\epsilon_{1}Y_{N_{1}}^{{\rm eq}}\left(0\right)\eta\left(K_{1},r_{1}\right),
\label{eq:approx_sol}
\end{eqnarray}
where $R\equiv Y_{N_{1}}^{{\rm eq}}\left(0\right)/Y^{{\rm eq}}=\frac{24}{\pi^{2}}$ and the efficiency factor is given by~\cite{Buchmuller:2004nz,Agashe:2018cuf}
\begin{eqnarray}
\eta\left(K_{i},r_{i}\right) & = & \begin{cases}
\frac{2}{Rb}e^{\frac{3\pi}{8}Rr_{i}K_{i}}\left\{ \exp\left[-\frac{\frac{3\pi}{8}Rr_{i}K_{i}}{\left(1+\sqrt{\frac{3\pi}{4}K_{i}}\right)^{2}}\right]-1\right\} \\
-\frac{2}{z_{B}Rr_{i}K_{i}}\left\{ 1-\exp\left[\frac{\frac{3\pi}{8}Rr_{i}K_{i}}{\left(1+\sqrt{\frac{3\pi}{4}K_{i}}\right)^{2}}z_{B}K_{i}\right]\right\}  & \mbox{for }Y_{N_{i}}\left(0\right)=0,\\
-\frac{2}{z_{B}Rr_{i}K_{i}}\left\{ 1-\exp\left[\frac{1}{2}z_{B}Rr_{i}K_{i}\right]\right\}  & \mbox{for }Y_{N_{i}}\left(0\right)=Y_{N_{i}}^{{\rm eq}}\left(0\right),
\end{cases}
\end{eqnarray}
with $K_{i}\equiv\frac{\Gamma_{N_{i}}}{{\cal H}\left(T=M_{i}\right)}$ and
\begin{eqnarray}
z_{B} & = & 1+\frac{1}{2}\ln\left[1+\frac{\pi R^{2}r_{i}^{2}K_{i}^{2}}{1024}\left(\ln\frac{3125\pi R^{2}r_{i}^{2}K_{i}^{2}}{1024}\right)^{5}\right].
\end{eqnarray}

\section{Casas-Ibarra parametrization with $\cpmutau$ symmetry}
\label{app:CI}

We review here the Casas-Ibarra parametrization in the presence of $\cpmutau$ symmetry.
It was used to produce the scatter plots in Figs.\,\ref{fig.1.f} and \ref{fig.3.f}.
Part of the formulas below were given in Ref.\,\cite{cp.mutau} with a slightly different notation.

Considering the seesaw formula in \eqref{Mnu:SS}, the Casa-Ibarra parametrization can be written as 
\eq{
\label{ap:mD}
m_D=\lambda v=i\hM_R^{1/2}R\hM_\nu^{1/2}U_\nu^\dag\,,
}
where $R$ is a complex orthogonal matrix that does not depend on low energy parameters.
The hatted matrices are the diagonalized matrices and $U_\nu$ is the PMNS matrix $V$.

The one-loop correction \eqref{Mnu:1-loop} to the neutrino mass matrix can be considered in the Casas-Ibarra parametrization by considering\,\cite{casas-ibarra:1-loop}
\eq{
\label{ap:mD:1-loop}
m_D^{\rm eff}=v\lambda^{\rm eff} =i\hM_R^{1/2}\big[1-C_{\rm eff}(\hM_R)\big]^{-1/2}R\hM_\nu^{1/2}U_\nu^\dag\,,
}
instead of \eqref{ap:mD}.

The $\cpmutau$ symmetry implies on $R$ the following symmetry\,\cite{cp.mutau}:
\eq{
\label{cpmutau:R}
R^*=-K_R^2 R K_\nu^2\,,
}
where $K_\nu^2$ and $(-K_R^2)$ contain the CP parities in the diagonal as one of
\eq{
(+++), (-++), (+-+), (++-)\,.
}
For example, the first option for $K_\nu^2$ means that $K_\nu^2=\id_3$.
We use the convention that $K_R$ and $K_\nu$ only contain 1 or $i$ in the diagonal.

The symmetry \eqref{cpmutau:R} implies that $R$ can be decomposed as
\eq{
\label{prop:R:corr}
R= -iK_R^* \Rz K_\nu\,,
}
with $\Rz$ being a real matrix obeying
\eq{
\label{prop:R:new:2}
{\Rz}^\tp (-K_R^2) \Rz = K_\nu^2\,,\quad
\Rz K_\nu^2 {\Rz}^\tp = -K_R^2\,.
}
When $K_\nu^2=\id_3$, also $(-K_R^2)=\id_3$, and $\Rz$ is a real orthogonal matrix in $O(3)$.
When $K_\nu^2=\diag(-++)$, Slvester's law tell us that $(-K_R^2)=\diag(-++)$ as well, except for reordering of the diagonal elements.
So ignoring the latter reordering, $\Rz$ is a member of the group $O(2,1)$, i.e., Lorentz transformations in 1+2 dimensions.

For the numerical sampling, we use the following parametrization:
\eqali{
K_\nu^2=(-K_R^2)=\id_3&:\quad \Rz=\pm\exp(A)\,,\quad 
A=\mtrx{0&\theta_3&-\theta_2\cr -\theta_3&0&\theta_1\cr \theta_2&-\theta_1&0}\,,
\cr
K_\nu^2=(-K_R^2)=\diag(-++)&:\quad \Rz=\pm\exp(A)\,,\quad 
A=\mtrx{0&\xi_1&\xi_2\cr \xi_1&0&\theta_1\cr \xi_2&-\theta_1&0}\,.
}
The other possibilities are obtained from permutations.
The angular variables are varied within
$\theta_i\in [-\pi,\pi]$ while the rapidity-like variables are varied on one of the following intervals:
\eq{
\xi_i\in [-3,3] \text{~~or~~}
\xi_i\in [-6,6]\,.
}
In Fig.\,\ref{fig.1.f} we have used the first interval.
In Fig.\,\ref{fig.3.f} we have used the first interval for the dark blue points and the second interval for the light blue points and red points.
To estimate the order of magnitude of the entries in $R$, recall that $\cosh(3)\approx\sinh(3)\approx 10$ and $\cosh(6)\approx\sinh(6)\approx 200$.

The PMNS matrix in the presence of $\cpmutau$ can be decomposed as\,\cite{cp.mutau,king.nishi}
\eq{
U_\nu=\Uz_\nu K_\nu\,,
}
where $\Uz_\nu$ obeys\,\cite{mutau-r:GL}
\eq{
(\Uz_\nu)_{ei}~\text{real and positive},
\quad
(\Uz_\nu)_{\mu i}=(\Uz_\nu)_{\tau i}^*\,,
}
$i=1,2,3$.
The rephasing freedom from the left can be fixed by choosing $\re(\Uz_\nu)_{\mu 3}=0$ and $\re(\Uz_\nu)_{\mu 2}>0$.
These properties fix $\theta_{23}=45^\circ$ and $\delta=\pm 90^\circ$.
For numerical calculations we choose $\delta=-90^\circ$ and use the 3$\sigma$ ranges for $\theta_{13}$ and $\theta_{12}$ in Ref.\,\cite{nufit}.
The lightest mass is restricted to $m_1\le 30\,\unit{meV}$ for NO and  
$m_3\le 16\,\unit{meV}$ for IO to respect the Planck limit of 120\,\unit{meV} for the sum of neutrino masses\,\cite{planck.2018}.
In Fig.\,\ref{fig.1.f} we have used a flat distribution for $m_0$ while 
in Fig.\,\ref{fig.3.f} we have used instead a flat distribution for $\log m_0$ with $m_0\ge 10^{-4}\,\unit{meV}$ in order to produce a larger number of points with small $m_0$.
The hierarchy of $M_i$ obeys $9M1\le 3M_2\le M_3$.


\end{document}